\documentclass{pasa}%

\usepackage{graphicx}
\usepackage{lscape}
\usepackage{arydshln}
\usepackage{soul}
\newcommand\arcmin{\hbox{$^\prime$}}
\newcommand\arcsec{\hbox{$^{\prime\prime}$}}
\newcommand\degree{\hbox{$^\circ$}}

\newcommand{\nisl}{{$\sim$ 2.3 million}}
\newcommand{\ncomp}{{$\sim$ 2.7 million}}
\newcommand{\skyarea}{30,480 }
\newcommand{\nislgalcut}{2,123,638}
\newcommand{\ncompgalcut}{2,462,693 }
\newcommand{\skyareagalcut}{28,020 }
\newcommand{\ntiles}{799 }

\newcommand{\flsumssN}{53,680 }
\newcommand{\flsumss}{1.00 }
\newcommand{\flsumssperr}{0.16 }
\newcommand{\flsumssnerr}{0.16 }

\newcommand{\rasumssN}{53,680 }
\newcommand{\rasumss}{+0.46 }
\newcommand{\rasumssperr}{4.05 }
\newcommand{\rasumssnerr}{3.66 }

\newcommand{\ranvss}{-0.71 }
\newcommand{\ranvssN}{286,735 }
\newcommand{\ranvssperr}{2.28 }
\newcommand{\ranvssnerr}{2.22 }

\newcommand{\raat}{-0.85 }
\newcommand{\raatN}{2505 }
\newcommand{\raatperr}{1.32 }
\newcommand{\raatnerr}{1.22 }

\newcommand{\decsumss}{+0.12 }
\newcommand{\decsumssperr}{2.51 }
\newcommand{\decsumssnerr}{2.62 }

\newcommand{\decnvss}{+0.31 }
\newcommand{\decnvssperr}{2.31 }
\newcommand{\decnvssnerr}{2.38 }

\newcommand{\decat}{+0.21 }
\newcommand{\decatperr}{0.77 }
\newcommand{\decatnerr}{0.86 }

\newcommand{\algleamN}{24,096 }
\newcommand{\algleam}{-0.69 }
\newcommand{\algleamperr}{0.25 }
\newcommand{\algleamnerr}{0.21 }

\newcommand{\alnvssN}{286,735 }
\newcommand{\alnvss}{-0.87 }
\newcommand{\alnvssperr}{0.52 }
\newcommand{\alnvssnerr}{0.42 }

\newcommand{\altgssN}{131,258 }
\newcommand{\altgss}{-0.64 }
\newcommand{\altgssperr}{0.26 }
\newcommand{\altgssnerr}{0.23 }

\newcommand{\altgssRN}{121,392 }
\newcommand{\altgssR}{-0.62 }
\newcommand{\altgssRperr}{0.25 }
\newcommand{\altgssRnerr}{0.22 }

\newcommand{\alfgleamN}{8,795 }
\newcommand{\alfgleam}{-0.61 }
\newcommand{\alfgleamperr}{0.32 }
\newcommand{\alfgleamnerr}{0.19 }

\newcommand{\alfnvssN}{222,436 }
\newcommand{\alfnvss}{-0.90 }
\newcommand{\alfnvssperr}{0.43 }
\newcommand{\alfnvssnerr}{0.39 }

\newcommand{\alftgssN}{45,512 }
\newcommand{\alftgss}{-0.59 }
\newcommand{\alftgssperr}{0.36 }
\newcommand{\alftgssnerr}{0.22 }

\newcommand{\alftgssRN}{41,989 }
\newcommand{\alftgssR}{-0.58 }
\newcommand{\alftgssRperr}{0.36 }
\newcommand{\alftgssRnerr}{0.21 }

\usepackage[dvipsnames]{xcolor}
\usepackage{pdflscape}
\usepackage{caption}
\usepackage{subcaption}

\title[RACS Stokes I Catalogue]{The Rapid ASKAP Continuum Survey Paper II: First  Stokes I Source Catalogue Data Release }
\author[C. L. Hale et al.]{C. L. Hale$^{1,2}$\thanks{Email: Catherine.Hale@ed.ac.uk}, D. McConnell$^3$, A. J. M. Thomson$^1$, E. Lenc$^3$, G. H. Heald$^1$,  A. W. Hotan$^1$, J. K. Leung$^{3,4}$, V. A. Moss$^3$, T. Murphy$^4$, J. Pritchard$^{4,3}$, E. M. Sadler$^{3,4}$, A. J. Stewart$^4$, M. T. Whiting$^3$
\affil{$^1$CSIRO Space and Astronomy, PO Box 1130, Bentley WA 6102, Australia}%
\affil{$^2$School of Physics and Astronomy, University of Edinburgh, Institute for Astronomy, Royal Observatory Edinburgh, Blackford Hill, Edinburgh EH9 3HJ, UK}%
\affil{$^3$CSIRO Space and Astronomy, PO Box 76, Epping, NSW, 1710, Australia}
\affil{$^4$Sydney Institute for Astronomy, School of Physics, University of Sydney, NSW 2006, Australia}
}%

\jid{PASA}
\doi{10.1017/pas.\the\year.xxx}
\jyear{\the\year}

\usepackage{aas_macros}
\usepackage{hyperref} 
\hypersetup{colorlinks,citecolor=blue,linkcolor=blue,urlcolor=blue}
\usepackage{float}

\begin{document}

\begin{frontmatter}
\maketitle

\begin{abstract}
The Rapid ASKAP Continuum Survey (RACS) is the first large sky survey using the Australian Square Kilometre Array Pathfinder (ASKAP), covering the sky south of +41\degree \ declination. With ASKAP's large, instantaneous field of view, $\sim 31$ deg$^2$, RACS observed the entire sky at a central frequency of {887.5 MHz using 903 individual pointings with 15 minute observations.} This has resulted in the deepest radio survey of the full Southern sky to date at these frequencies. In this paper, we present the first Stokes I catalogue derived from the RACS survey. This catalogue was assembled from \ntiles tiles that could be convolved to a common resolution of 25\arcsec, covering a large contiguous region in the declination range $\delta=-$80\degree \ to +30\degree. The catalogue provides an important tool for both the preparation of future ASKAP surveys and for scientific research. It consists of $\sim$2.1 million sources and excludes the $|b|<5$\degree \ region around the Galactic plane. This provides a first extragalactic catalogue with ASKAP covering the majority of the sky ($\delta<+30$\degree). We describe the methods to obtain this catalogue from the initial RACS observations and discuss the verification of the data, to highlight its quality. {Using simulations, we find this catalogue detects 95\% of point sources at an integrated flux density of $\sim$5 mJy. Assuming a typical sky source distribution model, this suggests an overall 95\% point source completeness at an integrated flux density $\sim$3 mJy.} The catalogue will be available through the CSIRO ASKAP Science Data Archive (CASDA).
\end{abstract}

\end{frontmatter}

\section{INTRODUCTION }
\label{sec:intro}

Radio surveys provide unique views into the Galactic and extragalactic skies. At the frequency {of the Rapid ASKAP Continuum {Survey} \citep[RACS, at 887.5 MHz;][]{RACS}}, and more generally below a few GHz, radio emission is dominated by synchrotron radiation; the emission from relativistic electrons spiralling within magnetic fields \citep{Condon1992}. This traces two main extragalactic populations: Active Galactic Nuclei (AGN) and Star Forming Galaxies (SFGs). For SFGs, it provides a method of obtaining unbiased star formation rates \citep[SFR; e.g.][]{Bell2003, Garn2009, Davies2017, Gurkan2018}, as radio emission is un-attenuated by dust. Observing synchrotron emission from AGN is important for understanding galaxy evolution, as their feedback is thought to limit the size to which galaxies can grow \citep[see e.g.][]{Bower2006, Fabian2012, Harrison2017}. Within the Galaxy, radio emission is often observed from supernova remnants \citep[see e.g.][]{Whiteoak1996, Anderson2017}, as Galactic synchrotron emission within the Galactic plane \citep[see e.g.][]{Haslam1982, MGPS, MGPS2, Wang2018} as well as from transient and variable sources \citep[see e.g.][]{Thyagarajan2011, Bhandari2018}. This variety of objects motivates radio surveys for advancing our understanding of the Universe. 

For catalogues of extragalactic radio sources, it is important to have both large area as well as deep observations. Deeper, smaller area surveys provide observations of fainter radio populations \citep[e.g. radio quiet quasars and SFGs, see e.g.][]{Wilman2008, Padovani2015, Smolcic2017b} and allow galaxy evolution to be investigated to earlier times in the age of the Universe. Large area surveys, on the other hand, allow extreme and rare AGN to be observed as well as large samples of resolved nearby SFGs. They are also crucial in providing information for radio sky models. Moreover, observations at multiple epochs of large sky areas are useful for detecting transient or variable sources {\citep[see e.g.][]{Thyagarajan2011, Mooley2016, Nyland2020}}.

At $\sim$1 GHz, radio surveys which have observed large regions of the southern skies ($\delta < 0$\degree) have been dominated by the combination of Sydney University Molonglo Sky Survey \citep[SUMSS;][]{SUMSS}, the Molongolo Galactic Plane Survey \citep[MGPS;][]{MGPS} and the updated MGPS-2 survey \citep{MGPS2} as well as the NRAO VLA Sky Survey \citep[NVSS;][]{NVSS}, complemented in the smaller overlap regions by Faint Images of the Radio Sky at Twenty-Centimeters \citep[FIRST;][]{FIRST, FIRST_2}. SUMSS surveyed the southern sky up to a northern-most $\delta =-$30\degree \ (excluding the Galactic plane $|b|<10$\degree) at 843 MHz with 45\arcsec/$\sin|\delta|$ resolution at a typical sensitivity of $\sim$1 mJy/beam. NVSS, on the other hand, is a northern sky survey which observed to a southern-most $\delta=-$40\degree \ at 1.4 GHz, observing with a constant 45\arcsec \ resolution at a typical sensitivity of $\sim$0.5 mJy/beam. FIRST also observed at 1.4 GHz with the VLA to a deeper sensitivity of $\sim$0.15 mJy/beam, at 5\arcsec \ resolution around the north and south Galactic caps. However, FIRST does not probe a large fraction of the southern skies and has limited sensitivity to extended emission. 

At lower frequencies, the Galactic and Extra-Galactic All-sky MWA Survey \citep[GLEAM;][]{GLEAM} and TIFR GMRT Sky Survey Alternate Data Release \citep[TGSS-ADR;][]{TGSS} provided observations of large regions of the southern sky. GLEAM observed south of $\delta$ = +30\degree \ at $\sim$2\arcmin \ resolution, reaching a root mean square (rms) sensitivity of 6$-$10 mJy/beam in the frequency range 70$-$230 MHz. TGSS-ADR observed the entire sky north of $\delta=-$53\degree \ at higher resolution, $\sim$25\arcsec, to an rms sensitivity of approximately 3.5 mJy/beam at 150 MHz. At higher frequencies, the Australia Telescope Compact Array (ATCA) has conducted surveys of the southern radio sky, including the Australia Telescope 20 GHz Survey \citep[AT20G;][]{AT20G}, with approximately 10\arcsec\ resolution, yielding a catalogue to an integrated flux density limit of 40 mJy (for $\delta<0$\degree). All these large area surveys are crucial to improving statistics on galaxy numbers, finding rare objects, as well as observing nearby radio sources and resolved star forming galaxies. Moreover, the combination of low, mid and high radio frequency radio surveys are important for spectral modelling of sources \citep[see e.g.][]{Clemens2008, Galvin2018}.

In order to observe such large areas it is advantageous to have an instrument with a large field of view. The Australian SKA Pathfinder \citep[ASKAP;][]{Johnston2008,Hotan2014, ASKAP} is one such facility able to easily conduct large sky surveys. It uses phased array feeds (PAFs), which provide an instantaneous field of view of 31 deg$^2$. The first large sky survey with ASKAP is the Rapid ASKAP Continuum Survey (RACS) and has been described in detail in Paper I \citep{RACS}. RACS used 15 minute observations to image the sky south of $\delta=+$41\degree \ using 903 tiles with typical sensitivities of 0.25$-$0.3 mJy/beam. Each tile is defined as the mosaic of the individual 36 beams which are simultaneously observed using the PAF technology. This survey {therefore provides} the best knowledge of radio sources at gigahertz frequencies in the southern hemisphere to date. In the future, the Evolutionary Map of the Universe {\citep[EMU;][]{Norris2011, EMU}} will provide a deeper map of the southern sky to $\sim20 \mu$Jy/beam rms, but this will require a significant increase in observation time. 

In this paper, we provide the first release of the RACS Stokes I source catalogue. The layout of this paper is as follows. Firstly, we present an overview of RACS in Section~\ref{sec:racs} and describe the observations used for this first data release. Next, we describe the cataloguing process in Section~\ref{sec:cat} and the final catalogue in Section~\ref{sec:fullcat}. We then discuss comparisons to previous surveys in Section~\ref{sec:previous_surveys} before discussing the completeness in Section~\ref{sec:comp_rel}. In Section~\ref{sec:sc}, we present the source counts, both raw and completeness-corrected, for this survey before drawing conclusions in Section~\ref{sec:conclusions}. 

\section{Rapid ASKAP Continuum Survey}
\label{sec:racs}

A detailed description of the RACS tiling and observation strategy can be found in Paper I \citep{RACS}. The majority of {RACS observations} were initially taken over the course of 12 days during April and May 2019. Subsequently, further re-observations of selected fields were taken in August 2019 and between March-June 2020. The latter re-observations were designed to reduce the PSF variation {amongst} the 36 beams within each individual tile. Each observation lasted 15 minutes in duration and a calibrator observation (of PKS\,B1934$-$638) of 200s in duration was typically observed within a day of each observation. These observations covered a 288 MHz bandwidth in the frequency range 744.5$-$1032.5 MHz. In total, 903 tiles were observed to ensure complete coverage of the sky for $\delta \leq$41$^{\circ}$.

Each observation was processed using \texttt{ASKAPSoft} \citep[see][ and Whiting et al. in prep]{Askapsoft1,Askapsoft2, Askapsoft3}. This software was specifically designed to take the raw ASKAP visibilities and produce calibrated images of the field, suitable for scientific use. The pipeline parameters used for the RACS survey are described thoroughly in Paper I. A robustness weight of 0.0 \citep[see][]{Briggs1995} was used, and short baselines were removed to improve image fidelity for those observations close to the Galactic plane (baselines smaller {than} 35~m {were} excluded) or affected by solar interference (baselines smaller than 100~m {were} excluded)\footnote{The baseline cuts imposed on each pointing can be found in the RACS database: \url{https://bitbucket.csiro.au/projects/ASKAP_SURVEYS/repos/racs/browse}}. All RACS tile images are on a 2.5\arcsec \ pixel grid and cover $\sim$31 deg$^2$. As described in Paper I, after the tile images {were} formed, a flux correction was applied to the tile to account for differences between the primary beam model applied within \texttt{ASKAPSoft} and the {response pattern} determined from holography measurements.

Images of each of the 903 tiles have been released publicly on the CSIRO ASKAP Science Data Archive {\citep[CASDA,][]{CASDA, CASDA2}}\footnote{{\url{https://data.csiro.au/}}}. To construct the first Stokes I catalogue we convolved each tile to a common resolution before mosaicing with neighbouring tiles to reduce sensitivity fluctuations.

\subsection{Selected Tiles}
\label{sec:selected}
A combination of factors affect the resolution of beam images within an individual tile, including: declination, hour angle coverage and the flagging of data within the observations. As ASKAP uses a phased array feed system, each field is constructed from 36 individual beams. The resolution can vary from beam to beam within an individual tile as well as between neighbouring tiles. In order to retain accurate flux scale measurements, it is necessary to ensure these beams and those in neighbouring tiles have the same resolution. This is {because} radio images are in units of Jy/beam and so varying PSFs in neighbouring images would affect both flux density and shape measurements of sources when mosaiced. 

\begin{figure*}
\begin{center}
\includegraphics[width=16cm]{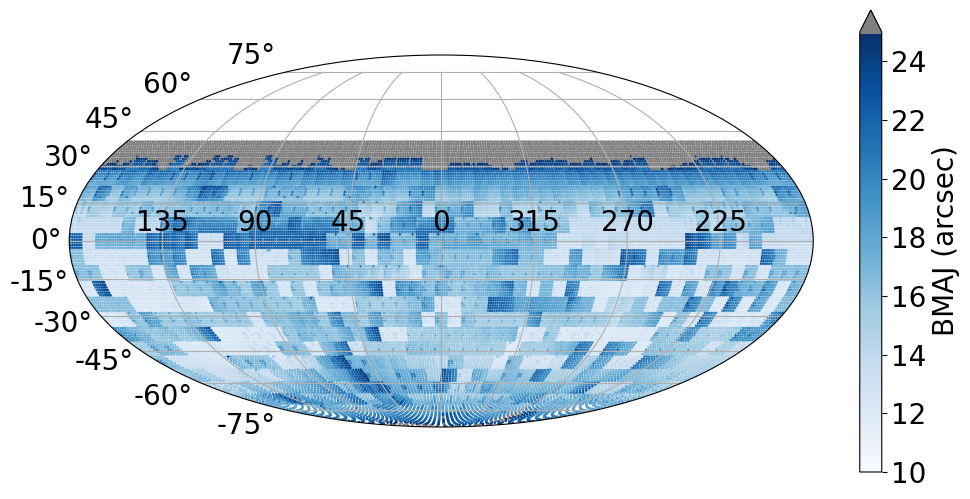} 
\caption{Sky variation of the PSF Major axis (FWHM) for all tiles selected in the database of \protect \cite{RACS} apart from those in Table \protect~\ref{tab:reobs} (see text). This is presented on an equatorial J2000.0 coordinate frame in Mollweide projection.} 
\label{fig:beamvariation}
\end{center}
\end{figure*}

To determine the fields to include in this first data release it {was} important to consider the aim of the survey. RACS will provide a model of the observable sky for future ASKAP surveys as well as provide an initial epoch as a benchmark for the search for variable or transient sources. For these purposes it is important to have a resolution as high as possible in order to resolve individual sources, and also to observe a large contiguous region of the sky.

However, it is not possible to {simply} use those individual beams which have angular resolution better than the desired criterion. This is as the holography corrections, as described in Paper I, requires all 36 beams of a tile to be present. This correction accounts for the differences between the beam models assumed in the \texttt{ASKAPSoft} linear mosaicing function, \texttt{linmos}, compared to that determined by holography measurements. As some beams within tiles can have poor angular resolution compared to other beams within the field, not all of the 903 RACS tiles {can} be convolved to a desired common resolution. 

Using these constraints, we decided upon a common resolution of a circular Gaussian beam with a diameter of 25\arcsec \ (Full Width at Half Maximum, FWHM) for the first catalogue data release. This choice of beam improves on the resolution of surveys such as SUMSS and NVSS by approximately a factor of 2, whilst also ensuring that the included observations cover the majority of the southern sky. The distribution of beam major axes (defined by the FWHM) across the entire RACS survey area released in Paper I is shown in Figure~\ref{fig:beamvariation}. All individual PSF major axes that are above 25\arcsec \ are shown in grey. For tiles in which there have been reobservations, the tile denoted with the column `SELECT=1' in the accompanying database to Paper I was generally chosen. However for a few fields, see Table~\ref{tab:reobs}, a different tile to the one selected in Paper I was used. This was because the tile selected could not be convolved to 25\arcsec, whilst a different observation of the same field could be. This often resulted in these fields having larger rms values, but did result in a continuous observing area. As illustrated in Figure~\ref{fig:beamvariation}, there can be large variations in the PSF major axis across the 903 tiles of the RACS observations.

\begin{table*}[]
    \centering
    \begin{tabular}{l l l l l l l}
        Tile ID & SBID & Median rms  & PSF$_0$ Major & SBID  & Median rms& PSF$_0$ Major \\
        & (Tile Used) & (Tile Used) & (Tile Used) & (Paper I) & (Paper I)  & (Paper I)\\
        & & (mJy/beam) & \arcsec & & (mJy/beam) & \arcsec \\ \hline \hline
        RACS\_0000+18A & 13780 & 0.70 & 16.7 & 8572 & 0.34 & 22.8 \\
        RACS\_0259-76A & 13715 & 0.54 & 19.6 & 9510 & 0.27 & 24.7  \\
        RACS\_0354-71A & 12534 & 0.26 & 19.7 & 8680 & 0.22 & 23.4 \\
        RACS\_1237+12A & 13586 & 1.03 & 17.2 & 10469 & 1.00 & 17.9 \\
        RACS\_1710+06A & 13761 & 0.84 & 14.1 & 8580 & 0.37 & 25.9 \\
        RACS\_2215+18A & 13707 & 0.63 & 16.9 & 8578 & 0.31 & 21.1 \\ \\
    \end{tabular}
    \caption{Table listing the six tiles for which this work used a different observation than the `best' observation in Paper I. For each tile we give scheduling block ID (SBID), median rms and PSF major axis of the reference beam.}
    \label{tab:reobs}
\end{table*}

\begin{figure*}
    \centering
    \includegraphics[width=12cm]{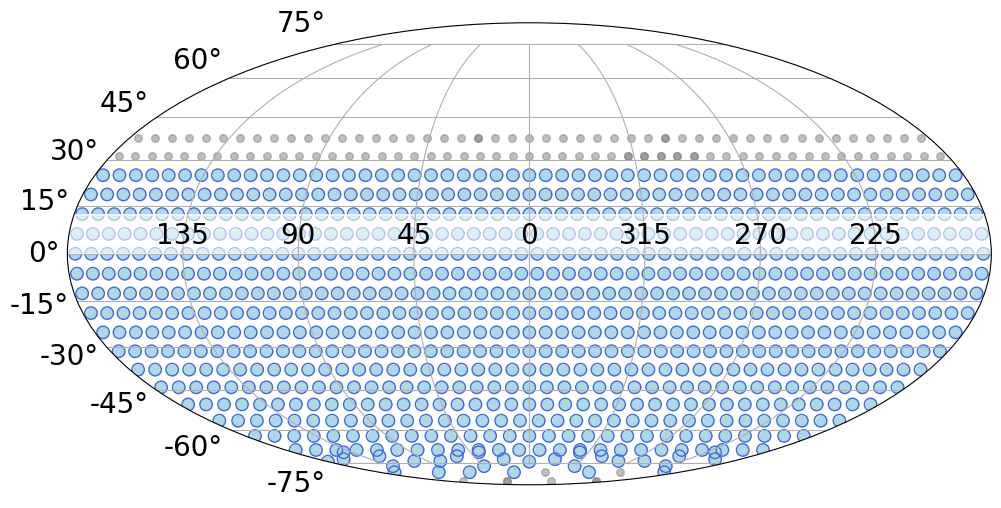}
    \caption{Coverage of tiles used in the first Stokes I catalogue for RACS. Fields that are within the coverage of this first catalogue release are shown in blue. Those fields within the RACS Paper I release that could not be convolved to 25\arcsec \ are shown in grey. In total \ntiles fields (of a possible 903) contribute to this work. This is presented on an equatorial J2000.0 coordinate frame in Mollweide projection.}
    \label{fig:RACS_coverage}
\end{figure*}

Figure~\ref{fig:RACS_coverage} shows the coverage for the first Stokes I catalogue release area across the sky compared to all of the RACS observations. The region for this first catalogue release (blue in Figure~\ref{fig:RACS_coverage}) covers the majority of the southern sky with $\delta=-$80\degree \ to +30\degree and compromises \ntiles fields (or tiles). 

\subsection{Producing Common Resolution Mosaics}
\label{sec:common_res}
In order to produce images at 25\arcsec \ resolution, we made use of scripts\footnote{using the \texttt{beamcon\_2D.py} function from \url{https://github.com/AlecThomson/RACS-tools}, version from 30$^{\textrm{th}}$ August 2020} to convolve each of the 36 single beam images to the desired 25\arcsec \ resolution, ensuring retention of the flux scale. This process is discussed further in \cite{RACS}. The convolved beam images were then linearly mosaiced together using the \texttt{ASKAPSoft} \texttt{linmos} function. Each beam image was weighted according to the number of contributing visibilities, and \texttt{linmos} assumed a circular Gaussian beam of FWHM $1.09 \lambda$/D for the primary beam model of each of the individual 36 beams. Here $\lambda$ is the central wavelength of the observations ($=c/\nu = c/887.5$ MHz $\sim$ 34cm) and D is the diameter of an ASKAP dish (12 m).

\begin{figure*}
    \begin{subfigure}{\textwidth}
    \begin{center}
    \centering
    \includegraphics[width=16cm]{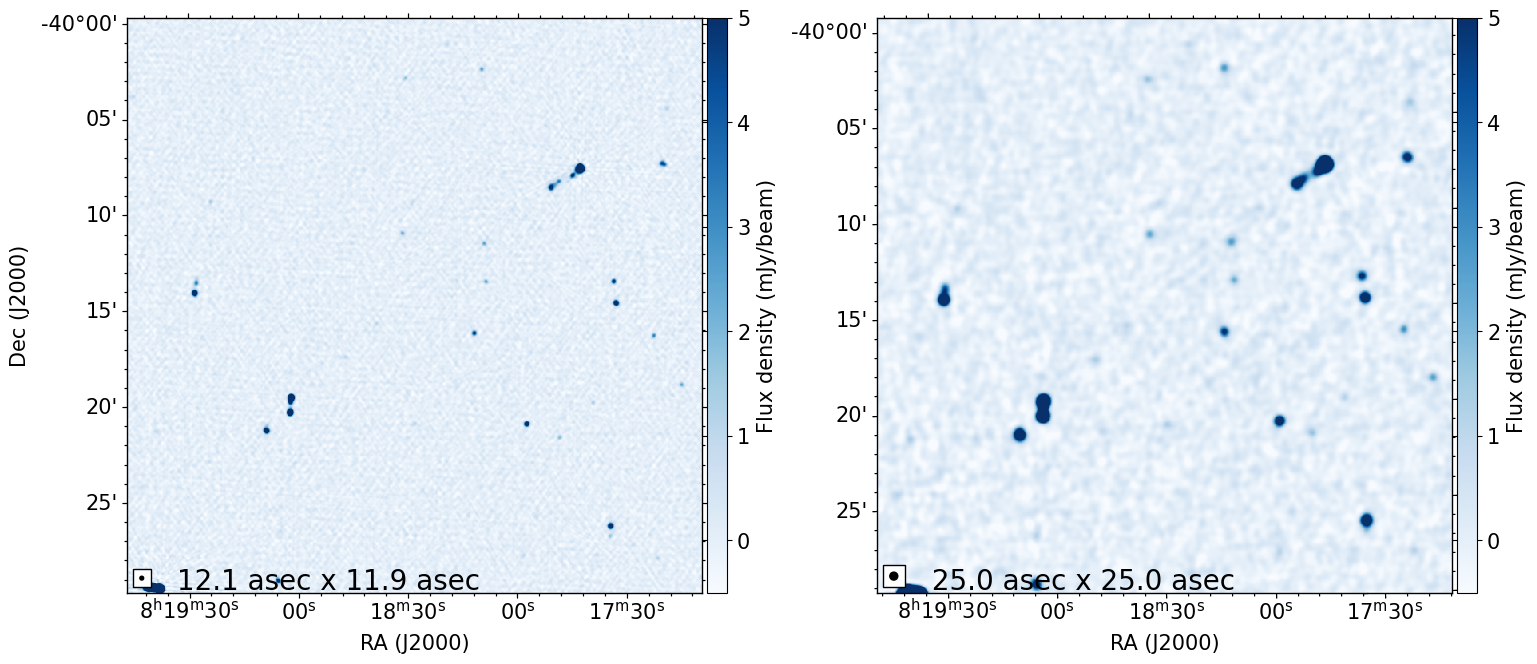}
    \subcaption{RACS\_0810-37A - Beam 27}
     \end{center}
    \end{subfigure}
    \begin{subfigure}{\textwidth}
    \begin{center}
    \centering
    \includegraphics[width=16cm]{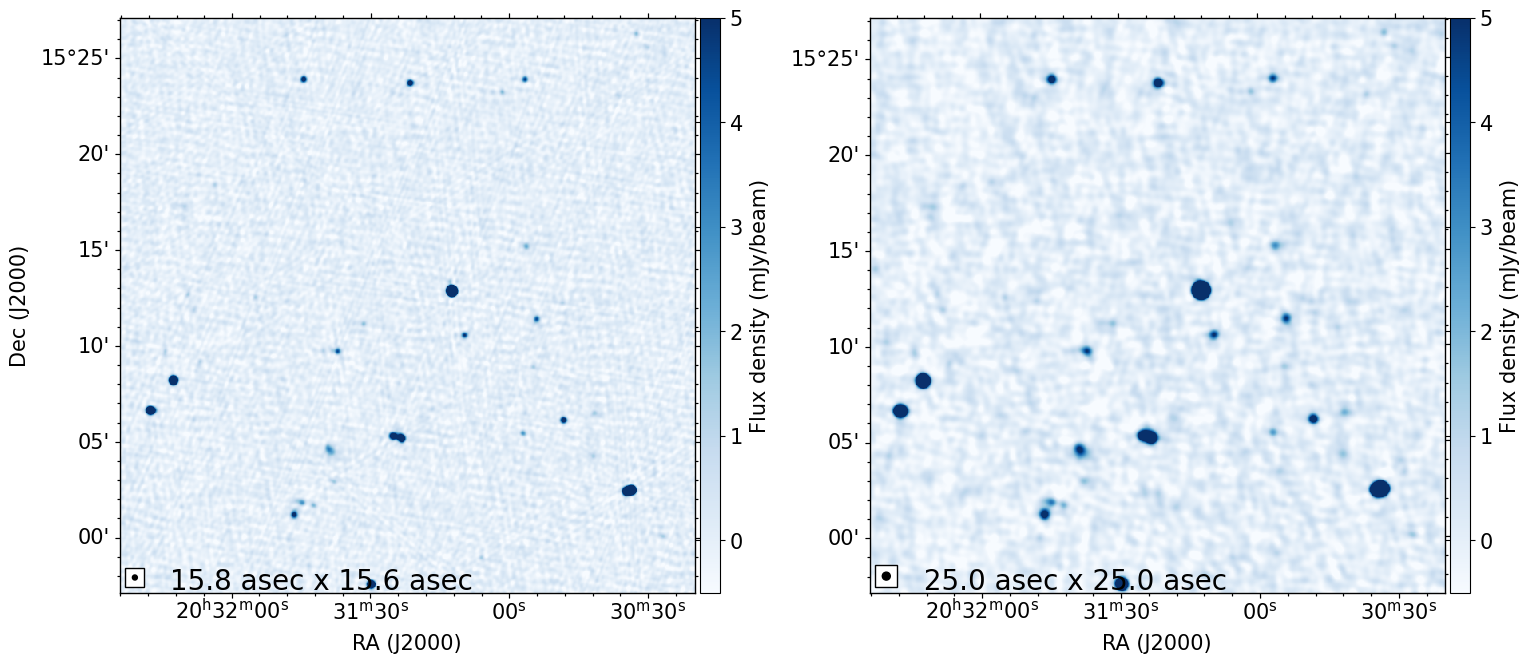}
     \subcaption{RACS\_2037+12A - Beam 17}
     \end{center}
    \end{subfigure}
    \begin{subfigure}{\textwidth}
    \begin{center}
    \centering
    \includegraphics[width=16cm]{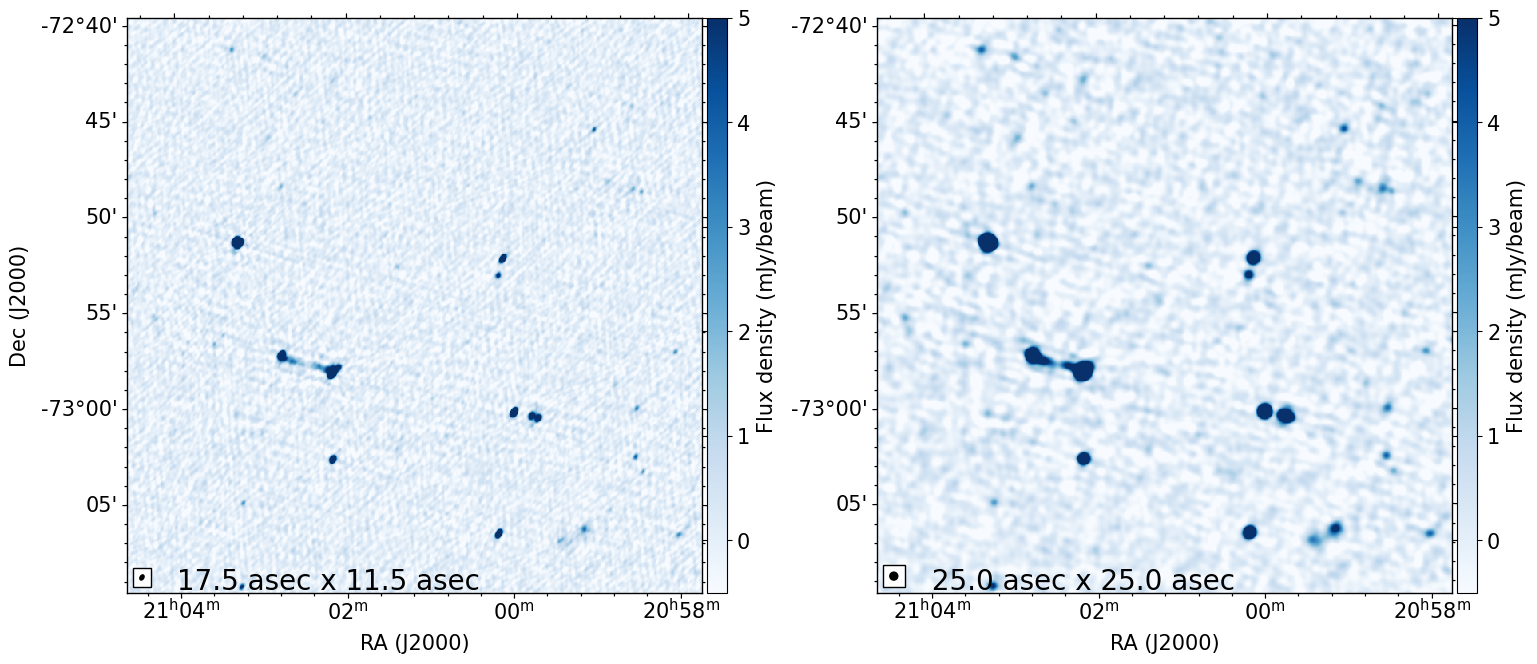}
     \subcaption{RACS\_2100-76A - Beam 16}
     \end{center}
    \end{subfigure}
    \caption{{Example comparison images of (left) the original beam image (pre-mosaicing) compared to (right)} those in the 25\arcsec \ mosaic tile image. This is shown for three fields with different angular resolutions (shown in the bottom left of each image). The colour scale for each image varies in the range $-$0.5 to 5 mJy/beam. Shown are tiles: RACS\_0810-37A, beam 27 (top); RACS\_2037+12A, beam 17 (middle); RACS\_2100-76A, beam 16 (bottom).}
    \label{fig:comp_25asec_images}
\end{figure*}

Figure~\ref{fig:comp_25asec_images} presents images before and after convolution to 25\arcsec \ resolution for three example regions, showing the differences in images for a range of pre-convolution major axis sizes. As can be seen, convolving to a poorer resolution loses some of the fine structure that could have been observed in the sources, such as in the jets of AGN, and has led to an apparent increase in the rms for each image. However, it does provide a consistent resolution across the full region used for this first Stokes I catalogue data release. This prioritises having a reliable flux scale across the image over {retaining} higher, but variable, angular resolution (which is still available for all tiles in CASDA). 

\subsection{Tile Flux Corrections}

As discussed in Paper I, the primary beam model assumed by \texttt{linmos} differs from the beam-dependent shapes revealed by holography measurements across the full ASKAP tile. This resulted in systematic and direction dependent errors in the brightness scale. Using a combination of holography and comparisons to SUMSS and NVSS, an empirical flux correction is applied to each tile. We apply this same flux correction to our linearly mosaiced, common resolution tiles. This flux correction varies across the field to a maximum correction of $\sim \pm30-40\%$. 

\subsection{Full image mosaic}

{We mosaiced} the convolved 25\arcsec, flux corrected tiles to improve the sensitivity at the edge of each tile. For each tile, adjacent tiles with overlapping regions were mosaiced together using \texttt{SWARP} \citep{Bertin2002} using the weights images produced with \texttt{linmos}. The resultant mosaiced tile image has the same extent as the original tile image but now includes contributions from neighbouring tiles. Since all tiles undergo the same mosaicing process, neighbouring mosaiced tiles will contain overlap regions with identical image data. 

These mosaiced tiles allow users to extract images over small specific regions with ease, compared to if a full mosaic of the entire sky existed. These mosaics as well as the source catalogue will be released on CASDA alongside the release of the paper\footnote{{The data will be available from \url{https://doi.org/10.25919/8zyw-5w85} alongside the release of this paper}}. Additionally, as a full image of the sky is important to be able to easily navigate the survey and search for objects, this is included in the form of a HiPS \citep{Fernique2015} image at \url{https://www.atnf.csiro.au/research/RACS/CRACS_I2/}. This HiPS image is an under-sampled version of the mosaiced images that are being released and allows a simple method for users to explore the entire RACS observations.  \\ 

\noindent These mosaiced images of the tiles form the basis of the Stokes I RACS catalogue.

\section{Stokes I Catalogue: Individual Tiles}
\label{sec:cat}

In order to generate the first full Stokes I catalogue of the area described in Section~\ref{sec:selected}, we first produced individual catalogues for each mosaiced tile. Further work was needed to combine these catalogues into a single Stokes I catalogue. In this Section, we describe the process of extracting the initial catalogues. The merger of the tile catalogs is then described in Section~\ref{sec:fullcat}.

We make use of the source extraction software \texttt{PyBDSF} \citep[using version 1.9.1,][]{PyBDSF}. \texttt{PyBDSF} was designed as a radio source finding tool for the LOw Frequency ARray \citep[LOFAR;][]{LOFAR} that identifies areas of radio emission (islands) and fits these regions with 2D elliptical Gaussian components in order to produce both a `Gaussian component' and `Source' catalogue. The `Gaussian component' catalogue (hereafter called component) consists of all the 2D Gaussians that are used to model sources in the field. As radio sources have a diverse range of morphologies, a combination of single and multiple component sources will exist within the catalogue. The source catalogue, in its default mode, joins together Gaussians within an island based on the separation of Gaussians and their flux values\footnote{see \url{https://www.astron.nl/citt/pybdsf/algorithms.html\#grouping-of-gaussians-into-sources}}. Because of this, Gaussians within the same island may be considered different sources. However, if necessary, it is also possible to force Gaussians of the same island to be grouped together in a single source. Details of \texttt{PyBDSF} and the parameters which users can specify can be found at \url{https://www.astron.nl/citt/pybdsf/}. 

When using \texttt{PyBDSF} on individual tiles, we specified several non-default parameters:

\begin{verbatim}
     - advanced_opts = True
     - thresh = `hard'
     - rms_box = (150,30)
     - atrous_do = True 
     - atrous_jmax = 3
     - mean_map = `zero'
     - frequency = 887.5e6
     - group_by_isl = True.
\end{verbatim}

By default, \texttt{PyBDSF} uses a 3$\sigma$ detection threshold to identify an island boundary (\texttt{thresh\_isl = 3}), and a 5$\sigma$ threshold is used to include islands within a catalogue (\texttt{thresh\_pix = 5}). Setting \texttt{thresh = `hard'} enforces this 5$\sigma$ cut, and does not include a variable threshold based on the false detection rate.

We also specify the box size used by \texttt{PyBDSF} to generate an rms map through specifying \texttt{rms\_box = (150,30)}. Whilst \texttt{PyBDSF} can internally determine an appropriate size of box in order to produce the rms image, this may need to be changed if there are artefacts within the image\footnote{see \url{https://www.astron.nl/citt/pybdsf/examples.html\#image-with-artifacts}}. In fact, when we did not specify this parameter, the box size determined by \texttt{PyBDSF} could be as large as approximately 1000 pixels across. This was found to be too large and artefacts around bright sources were being included by \texttt{PyBDSF} in the output catalogue produced. Therefore we decided to specify a smaller box size to better account for bright sources and remove the likelihood of artefacts being confused for real sources. The 150 in the rms box size represents the box size used to calculate the rms. It was chosen to be 150 pixels as this appeared to reflect the scale {over} which artefacts influenced the image surrounding a bright source for the areas with artefacts investigated. The 30 in \texttt{rms\_box} reflects the step size (in pixels) by which the box is moved to calculate the rms.

Moreover, because of the sensitivity of ASKAP to extended emission, we wanted to ensure {that such sources} were accurately modelled by \texttt{PyBDSF}. To do this we followed advice from the \texttt{PyBDSF} pages\footnote{see \url{https://www.astron.nl/citt/pybdsf/examples.html\#image-with-extended-emission}}. We set the \texttt{mean\_map} parameter to `zero' and switched on the atrous mode using \texttt{atrous\_do = True}. We used fitting up to three wavelet scales in this mode through setting \texttt{atrous\_jmax = 3}. This allows extended emission on larger scales to be fit. As the rms appeared to vary across the field especially around bright sources, we left the \texttt{rms\_map} as the default parameter in which an rms map is calculated for the field using the rms box size specified.

Finally, due to the source density within these observations, we believe we are not limited by confusion (see Section~\ref{sec:galcat} for more details on the source density). By setting \texttt{group\_by\_isl = True}, we made the assumption that all sources within the same island are likely associated with the same source. From visually investigating a handful of random fields, the models produced by \texttt{PyBDSF} seemed to model source emission of resolved sources well. 

After running \texttt{PyBDSF} we recorded {three} things: 
\begin{itemize}
    \item The catalogue of Gaussians identified within the image
    \item The catalogue of grouped sources identified within the image
    \item An rms image of the field
\end{itemize}

Both the `Gaussian' and `Source' catalogues have scientific value. The Gaussian catalogue is useful as it can be used to de-blend the emission from close neighbouring sources which are not associated with one another. However, the `Source' catalogue is useful for providing information on multi-component sources. Therefore we construct and release both a `Gaussian' and `Source' catalogue associated with the data. 

\section{Full Sky Catalogue}
\label{sec:fullcat}

We compose the catalogue from the PyBDSF outputs giving each entry a unique identifier by combining field name and PyDBSF source/component identifiers. For example, source 0 in tile \texttt{RACS\_0000+12A} was renamed from a \texttt{Src\_ID} of 0 to \texttt{RACS\_0000+12A\_0}. An extra column that included the \texttt{Tile\_ID} associated with a source and its separation from the tile centre was also recorded.

Due to the overlapping tiles, a simple concatenation of all the individual catalogues would result in the duplication of sources. As the images within the overlap regions are identical, only sources detected in a given tile for which that tile centre is the closest to the source are included in the final catalogue. The source position, not the position of individual components, is used for this match. This is due to the possibility of different Gaussians within the same source near a tile boundary having different tile centres as their nearest tile. After concatenation, we ensure that no sources from different tiles were separated by less than 2 pixels (i.e. 5\arcsec).  This only affected a very small number of sources (3 pairs - i.e. 6 sources), and so duplicates of these were removed.

We {rounded the data to a given number of decimal places for the column} also apply another 5$\sigma$ thresholding. Whilst \texttt{PyBDSF} uses a 5$\sigma$ threshold, this will be based on the peak pixel value within the image, not the modelled peak flux. This can therefore be greatly affected by noise fluctuations. To ensure we have high SNR sources we therefore ensure a 5$\sigma$ cut using the peak flux recorded in the \texttt{PyBDSF} catalogue and the island rms column. 

\begin{figure*}
\begin{center}
\includegraphics[width=16cm]{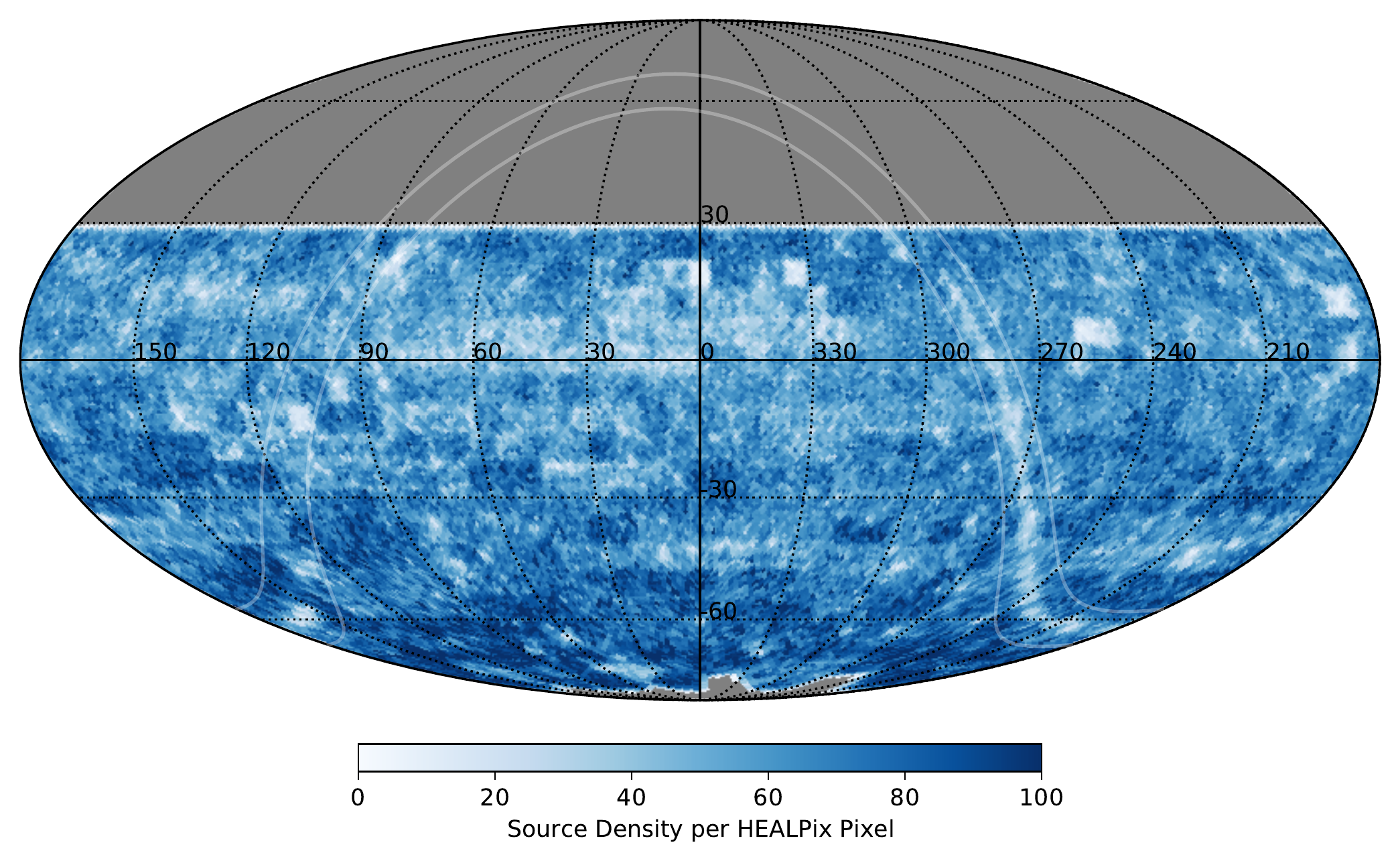}
\caption{Number density of sources per HEALPix pixel (each approximately 55\arcmin \ in size) across the sky from the merged sky catalogue, including the Galactic plane. The Galactic plane region  ($|b|=5$\degree) is indicated by the faint solid white lines. This is presented on an equatorial J2000.0 coordinate frame in Mollweide projection.} 
\label{fig:sky_dist}
\end{center}
\end{figure*}

Combining components and sources in this manner produced an initial source catalogue over the majority of the southern sky ($\delta=-80$\degree \ to $+30$\degree) of {\nisl} \ radio sources and a corresponding component catalogue of {\ncomp} \ components, covering a total sky area of \skyarea deg$^2$. Figure~\ref{fig:sky_dist} presents the observed density of sources across the sky using a HEALPix grid\footnote{\url{http://healpix.sf.net}; using the healpy python module \citep{Gorski2005, Zonca2019}}, with an N$_{\rm side}$ value of 64, corresponding to a rough pixel size of 55\arcmin. The apparent source density variation is discussed later in this paper.

\subsection{Noise Distribution}
\begin{figure*}
\begin{center}
\includegraphics[width=16cm]{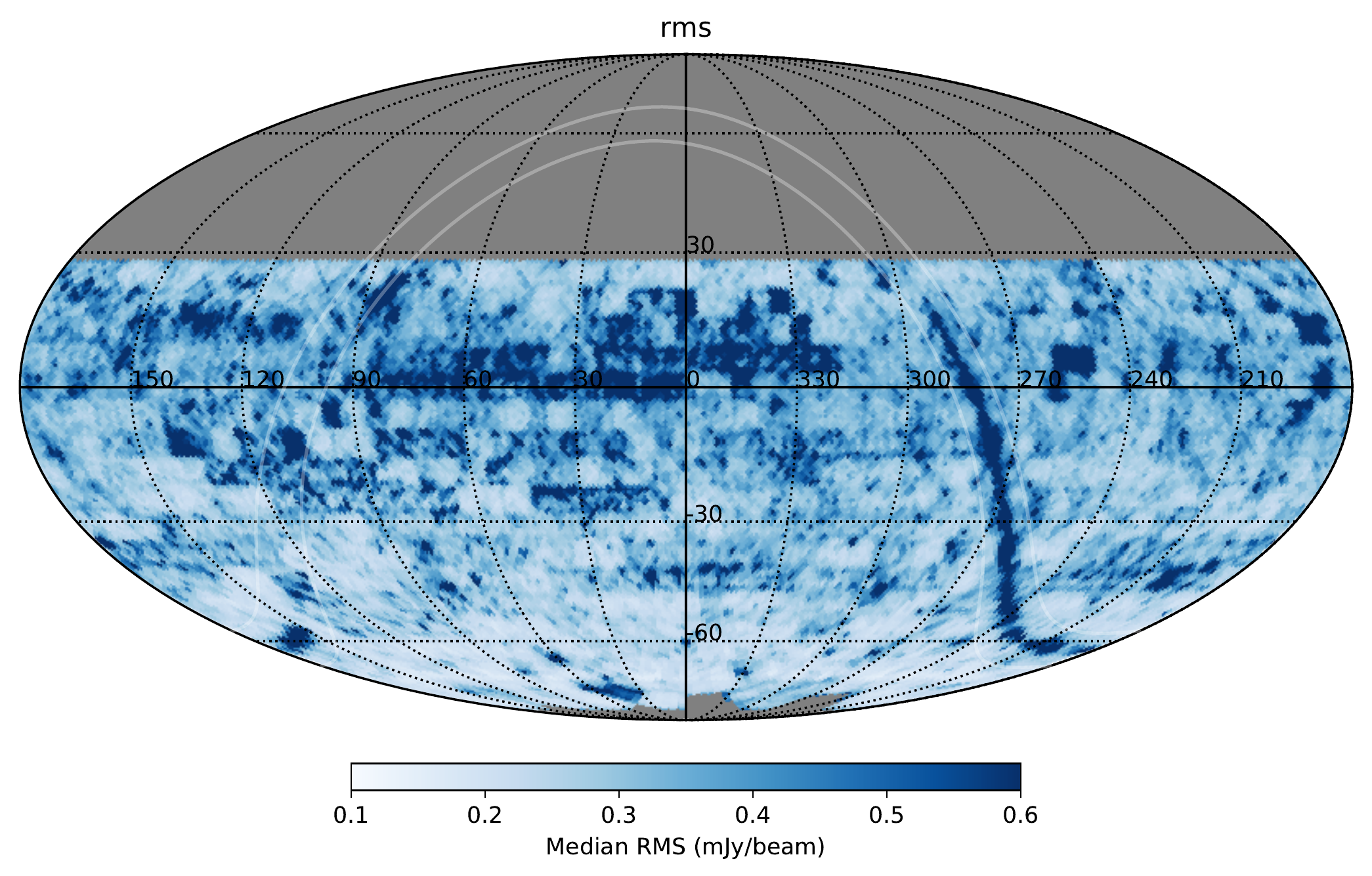} 
\caption{{rms distribution across the sky based on selecting random positions across the survey region and calculating the median rms in each HEALPix bin. The Galactic plane ($|b|=5$\degree) is indicated by the faint solid white lines. This is presented on an equatorial J2000.0 coordinate frame in Mollweide projection. } }
\label{fig:rms_dist}
\end{center}
\end{figure*}

When \texttt{PyBDSF} produces the source and Gaussian catalogue of each tile, a variable rms map of each image is generated. In order to present this rms variation across this first data release, we randomly select 10 million positions across the sky in the range $\delta= -$85\degree \ to +30\degree. At each position, the value of the rms at that location is recorded. We plot this distribution of rms in Figure~\ref{fig:rms_dist} on the same HEALPix grid as above and plotting the median rms value within each HEALPix cell. 

The rms varies across the RACS survey due to a combination of factors. This includes the proximity to bright sources, unmodelled extended emission (which may be a factor close to the Galactic plane), conditions such as the hour angle coverage of the observations and the overlap of tiles across the sky.
{As shown in Figure~\ref{fig:rms_dist}, there are large rms values along the Galactic plane as well as in other regions for example around $\delta=0$\degree.} These variations across the full sky will arise from a variety of reasons such as from having extended emission in the Galactic plane; having bright sources with large artefacts within a field and, finally, the scheduling of the observations relative to its hour angle coverage.
The median rms is typically smaller at more southerly declinations compared to equatorial regions. This may be influenced by the greater overlap between neighbouring tiles or possibly due to the hour angle coverage of these observations (see Paper I) 

\begin{figure*}
\begin{subfigure}{0.5\textwidth}
\includegraphics[width=8cm]{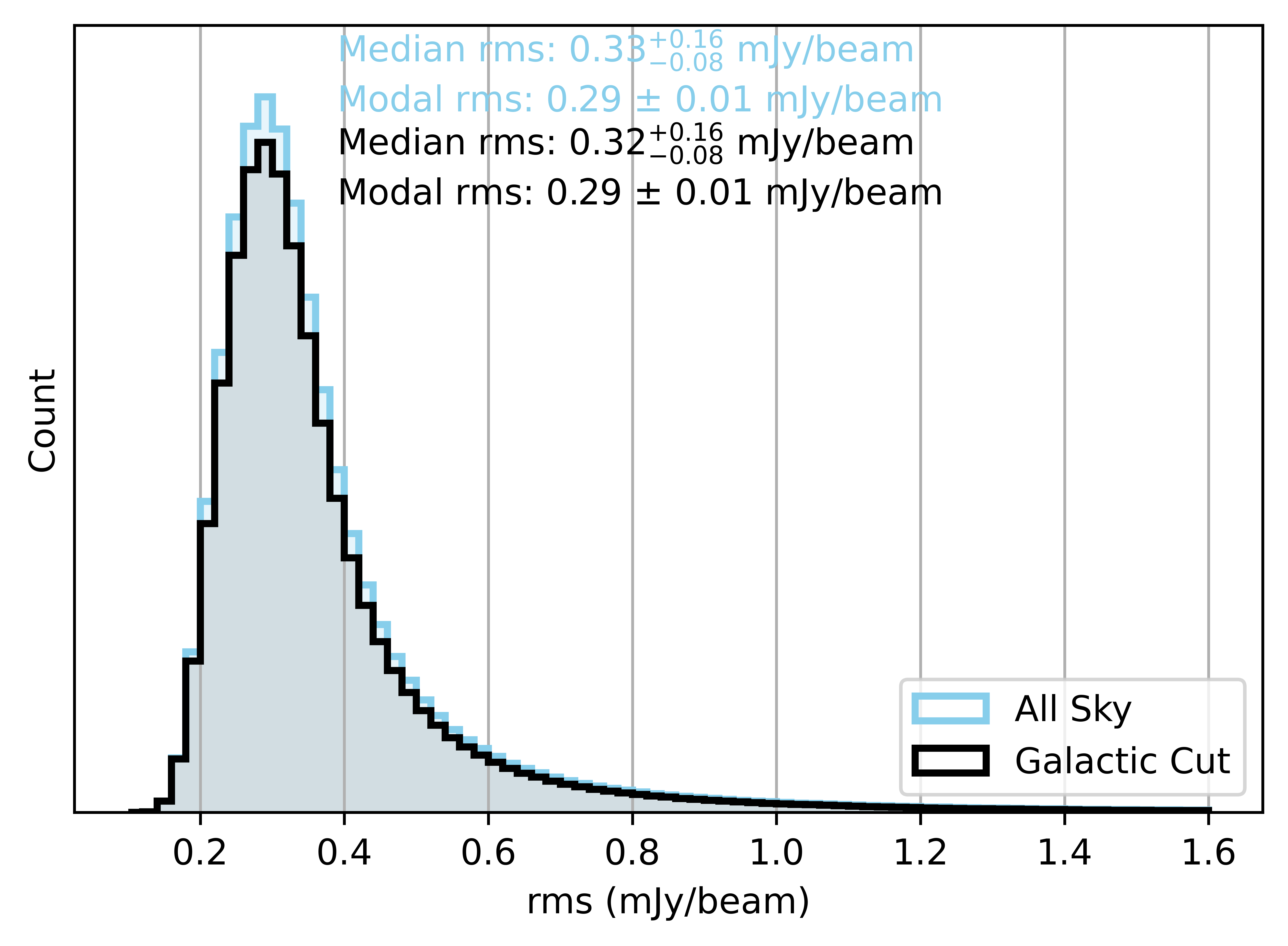}
\subcaption{}
\end{subfigure}
\begin{subfigure}{0.5\textwidth}
\includegraphics[width=8cm]{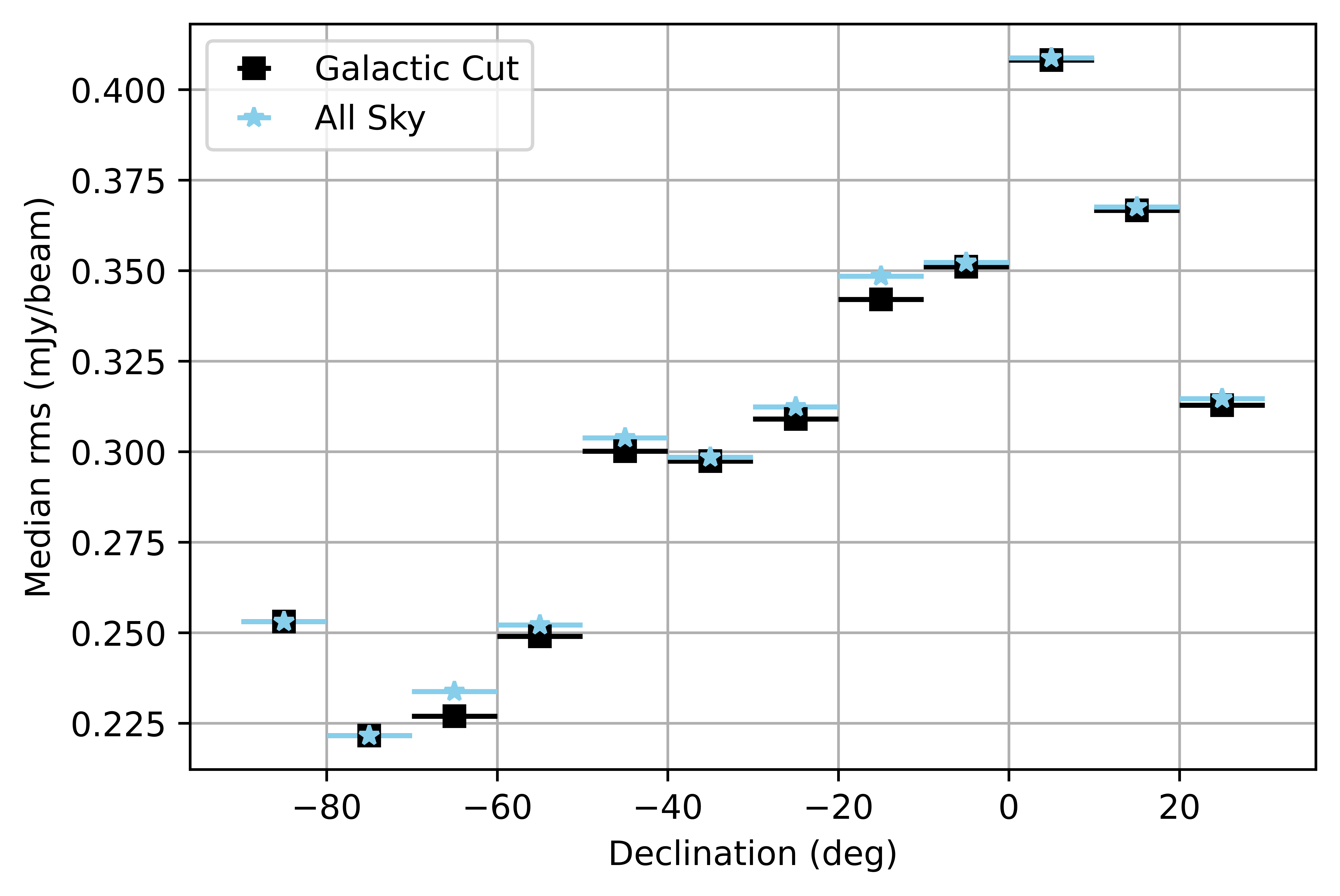}
\subcaption{}
\end{subfigure}
\caption{{Left: Histogram of the rms distribution from randomly selecting positions across the sky. Right: The median rms of each declination strip as a function of declination with (blue) and without ({black}) the Galactic plane included.} }
\label{fig:rms}
\end{figure*}

The distribution of all rms values (from these random positions) across the field of view (\skyarea \ deg$^{2}$) can be seen in Figure~\ref{fig:rms} (left), and the variation of the median rms value as a function {of} declination within different declination bins can be seen in Figure~\ref{fig:rms} (right). This is shown
{both inclusive and exclusive of the Galactic plane.}
Across the full sky, the rms values typically have a median value of approximately 0.3 mJy/beam, however this is closer to $0.2-0.25$ mJy/beam at $\delta \lesssim -50$\degree \ rising to values closer to $0.35-0.4$ mJy/beam near $\delta=$ 0\degree.

\subsection{The Galactic Plane}
\label{sec:galcat}
As can be seen in Figure~\ref{fig:rms_dist}, the rms is elevated around the Galactic plane. Furthermore, as presented in Paper I, the emission around the Galactic plane includes substantial extended emission, such as supernova remnants. As these structures will be insufficiently modelled using Gaussian components, we removed the region around the Galactic plane. Therefore whilst the images on CASDA will contain these regions, the final catalogues {used for the analysis in this paper} do not contain any sources where the magnitude of the Galactic latitude, $|b| <5$\degree. Also, in regions near the Large and Small Magellanic Clouds or supernova remnants sources may be poorly modelled as Gaussian components, however these regions remain in the catalogue. 

After excluding the low Galactic latitudes the final source catalogue contains {\nislgalcut} sources and {\ncompgalcut}  Gaussian components over $\sim$\skyareagalcut sq. deg of the sky. This corresponds to an average $\sim$90 components or $\sim$75 sources per square degree.

{We include with the data release the catalogue generated within the galactic plane region defined here. However, we urge caution for any users wanting to use this catalogue for regions with $|b| <5$\degree. All further quality assessment and comparison of the catalogue to previous surveys refers solely to the catalogue with the galactic latitude cut imposed and we note that, as can be seen in \cite{RACS}, there may be flux density offsets close to the galactic field, as well as large RMS values (see Figure \ref{fig:rms}).}

\subsection{Catalogue Columns}
\label{sec:fincat}
Using the combined \texttt{PyBDSF} catalogues, we use a subset of the column information when generating the final catalogues. These columns provide information on: IDs; astrometry; flux densities; shape information and other important source information. 
We present an example of the first 10 lines of the source catalogue in Table~\ref{tab:islcat} and {Gaussian} component catalogue in Table~\ref{tab:compcat}, sorted by the Source\_ID of the source/{Gaussian} component. We present a description of the column information for these two RACS Stokes I catalogues below\footnote{{We note that on formatting this catalogue for release we rounded the columns to an appropriate number of decimal places. Upon doing this 11 {Gaussian} components had integrated flux densities $<$ 0.001mJy and so remain in the catalogue with {Gaussian} component flux density = 0 mJy.}}.

\begin{table*}
\begin{minipage}{\textwidth}
\begin{center}
\centering
\begin{tabular}{lrrrr} 
\\ \hline
Source\_Name & Source\_ID & Tile\_ID & SBID & Obs\_Start\_Time \\ 
  &  &  &  &  \\ \hline \hline 
 RACS-DR1 J001237.8+101809 & RACS\_0000+12A\_1071 & RACS\_0000+12A & 8578 & 58599.06929 \\  
 RACS-DR1 J001237.3+101939 & RACS\_0000+12A\_1072 & RACS\_0000+12A & 8578 & 58599.06929 \\  
 RACS-DR1 J001236.9+110238 & RACS\_0000+12A\_1079 & RACS\_0000+12A & 8578 & 58599.06929 \\  
 RACS-DR1 J001234.3+110628 & RACS\_0000+12A\_1084 & RACS\_0000+12A & 8578 & 58599.06929 \\  
 RACS-DR1 J001231.2+103907 & RACS\_0000+12A\_1088 & RACS\_0000+12A & 8578 & 58599.06929 \\  
 RACS-DR1 J001232.9+114652 & RACS\_0000+12A\_1089 & RACS\_0000+12A & 8578 & 58599.06929 \\  
 RACS-DR1 J001234.1+124520 & RACS\_0000+12A\_1090 & RACS\_0000+12A & 8578 & 58599.06929 \\  
 RACS-DR1 J001235.3+134158 & RACS\_0000+12A\_1092 & RACS\_0000+12A & 8578 & 58599.06929 \\  
 RACS-DR1 J001228.3+112827 & RACS\_0000+12A\_1098 & RACS\_0000+12A & 8578 & 58599.06929 \\  
 RACS-DR1 J001226.8+105716 & RACS\_0000+12A\_1100 & RACS\_0000+12A & 8578 & 58599.06929 \\ \hline 
  \end{tabular} 
 \end{center}
 \end{minipage} \\ 

\begin{minipage}{\textwidth}
\begin{center}
\centering
\begin{tabular}{rrrrrrrrr} 
\\ \hline
N\_Gaus & RA & Dec & E\_RA & E\_DEC & Total\_flux & E\_Total\_flux & E\_Total\_flux & Peak\_flux \\ 
 &  &  &  &  & \_Source & \_Source\_PyBDSF & \_Source &  \\ 
  & \degree & \degree & \arcsec & \arcsec & mJy & mJy & mJy & mJy/beam \\ \hline \hline 
 1 & 3.157688 & 10.302451 & 1.75 & 1.04 & 3.560 & 0.666 & 1.003 & 2.968 \\  
 1 & 3.155227 & 10.327487 & 0.24 & 0.23 & 14.890 & 0.562 & 1.642 & 14.881 \\  
 1 & 3.153751 & 11.043854 & 3.45 & 4.50 & 7.158 & 1.613 & 1.898 & 2.060 \\  
 1 & 3.142925 & 11.107845 & 1.10 & 1.26 & 3.163 & 0.628 & 0.957 & 3.318 \\  
 1 & 3.129824 & 10.652017 & 1.67 & 2.52 & 2.563 & 0.695 & 0.972 & 1.971 \\  
 1 & 3.137132 & 11.781003 & 1.71 & 1.32 & 4.822 & 0.889 & 1.222 & 3.460 \\  
 1 & 3.142223 & 12.755545 & 2.17 & 1.26 & 13.814 & 1.632 & 2.195 & 5.772 \\  
 1 & 3.147178 & 13.699465 & 0.05 & 0.05 & 73.474 & 0.572 & 5.672 & 71.689 \\  
 1 & 3.117901 & 11.474149 & 0.11 & 0.11 & 41.977 & 0.760 & 3.521 & 40.863 \\  
 1 & 3.111569 & 10.954526 & 2.36 & 2.15 & 2.343 & 0.752 & 1.003 & 2.020 \\   \hline 
 \end{tabular} 
 \end{center}
\end{minipage}\\ 

\begin{minipage}{\textwidth}
\begin{center}
\centering
\begin{tabular}{rrrrrrrrrr} 
\\ \hline
E\_Peak\_flux & Maj & E\_Maj & Min & E\_Min & PA & E\_PA & DC\_Maj & E\_DC\_Maj & DC\_Min \\ 
 mJy/beam & \arcsec & \arcsec & \arcsec & \arcsec & \degree & \degree & \arcsec & \arcsec & \arcsec \\ \hline \hline  
 0.336 & 32.24 & 4.23 & 23.31 & 2.27 & 105.09 & 17.11 & 0.00 & 4.23 & 0.00 \\  
 0.324 & 25.44 & 0.56 & 24.64 & 0.53 & 80.74 & 29.91 & 0.00 & 0.56 & 0.00 \\  
 0.368 & 60.15 &12.06& 36.16 & 5.76 & 33.30 & 20.82 & 54.70 & 12.06 & 26.12 \\  
 0.374 & 25.58 & 3.03 & 23.33 & 2.50 & 158.81 & 51.49 & 0.00 & 3.03 & 0.00 \\  
 0.334 & 31.86 & 5.95 & 25.56 & 3.92 & 0.26 & 35.41 & 19.73 & 5.95 & 5.13 \\  
 0.407 & 32.56 & 4.16 & 26.79 & 2.91 & 111.73 & 27.96 & 20.82 & 4.16 & 9.61 \\  
 0.495 & 53.55 & 5.57 & 27.97 & 1.98 & 116.00 & 8.56 & 47.34 & 5.57 & 12.54 \\  
 0.325 & 25.49 & 0.12 & 25.17 & 0.11 & 100.85 & 17.02 & 4.79 & 0.12 & 2.89 \\  
 0.431 & 25.47 & 0.27 & 25.25 & 0.26 & 154.41 & 43.48 & 4.85 & 0.27 & 3.27 \\  
 0.392 & 28.78 & 5.94 & 25.24 & 4.60 & 124.68 & 64.18 & 14.23 & 5.94 & 3.17 \\   \hline 
 \end{tabular} 
 \end{center}
\end{minipage}\\ 

\begin{minipage}{\textwidth}
\begin{center}
\centering
\begin{tabular}{rrrrrrrrr} 
\\ \hline
E\_DC & DC\_PA & E\_DC\_PA & S\_Code & Separation & Noise & Gal\_lon & Gal\_lat & Flag \\ 
\_Min& &  &  & \_Tile\_Centre &  & & & \_Close\\ 
 \arcsec & \degree & \degree &  & \degree & mJy/beam & \degree & \degree &  \\ \hline \hline 
 2.27 &    0.00 & 17.11 & S & 3.8431 & 0.329 & 107.518268 & -51.404165 & -  \\  
 0.53 &    0.00 & 29.91 & S & 3.8262 & 0.324 & 107.524147 & -51.379281 & -  \\  
 5.76 &  33.30 & 20.82 & S & 3.4484 & 0.367 & 107.793667 & -50.683253 & -  \\  
 2.50 &    0.00 & 51.49 & S & 3.4105 & 0.379 & 107.801283 & -50.618566 & -  \\  
 3.92 &    0.26 & 35.41 & S & 3.6220 & 0.319 & 107.609691 & -51.058232 & -  \\  
 2.91 & 111.73 & 27.96 & S & 3.1691 & 0.385 & 108.040448 & -49.962954 & -  \\  
 1.98 & 116.00 & 8.56 & S & 3.0707 & 0.488   & 108.394965 & -49.015996 & -  \\  
 0.11 & 100.85 & 17.02 & S & 3.2624 & 0.323 & 108.725686 & -48.097792 & -  \\  
 0.26 & 154.41 & 43.48 & S & 3.2441 & 0.428 & 107.899652 & -50.256826 & -  \\  
 4.60 & 124.68 & 64.18 & S & 3.4531 & 0.380 & 107.696621 & -50.760190 & -  \\  \hline  
\end{tabular} 
\end{center}
\caption{{The first 10 lines from the final source catalogue. The columns are described in Section~\ref{sec:srccat}.}}
\label{tab:islcat}
\end{minipage}
\end{table*} 

\begin{table*} 
\begin{minipage}{\textwidth}
\begin{center}
\centering
\small
\begin{tabular}{rrrrrrr} 
\\ \hline
Gaussian\_ID & Source\_ID & Tile\_ID & SBID & Obs\_Start\_Time & N\_Gaus  & RA \\ 
  &  &  &  &  & &\degree  \\ \hline \hline 
 RACS\_0000+12A\_1243 & RACS\_0000+12A\_1071 & RACS\_0000+12A & 8578 & 58599.06929 & 1 & 3.157688 \\  
 RACS\_0000+12A\_1244 & RACS\_0000+12A\_1072 & RACS\_0000+12A & 8578 & 58599.06929 & 1 & 3.155227 \\  
 RACS\_0000+12A\_1251 & RACS\_0000+12A\_1079 & RACS\_0000+12A & 8578 & 58599.06929 & 1 & 3.153751 \\  
 RACS\_0000+12A\_1256 & RACS\_0000+12A\_1084 & RACS\_0000+12A & 8578 & 58599.06929 & 1 & 3.142925 \\  
 RACS\_0000+12A\_1260 & RACS\_0000+12A\_1088 & RACS\_0000+12A & 8578 & 58599.06929 & 1 & 3.129824 \\  
 RACS\_0000+12A\_1261 & RACS\_0000+12A\_1089 & RACS\_0000+12A & 8578 & 58599.06929 & 1 & 3.137132 \\  
 RACS\_0000+12A\_1262 & RACS\_0000+12A\_1090 & RACS\_0000+12A & 8578 & 58599.06929 & 1 & 3.142223 \\  
 RACS\_0000+12A\_1265 & RACS\_0000+12A\_1092 & RACS\_0000+12A & 8578 & 58599.06929 & 1 & 3.147178 \\  
 RACS\_0000+12A\_1272 & RACS\_0000+12A\_1098 & RACS\_0000+12A & 8578 & 58599.06929 & 1 & 3.117901 \\  
 RACS\_0000+12A\_1274 & RACS\_0000+12A\_1100 & RACS\_0000+12A & 8578 & 58599.06929 & 1 & 3.111569 \\ \hline 
 \end{tabular}
 \end{center}
\end{minipage} \\ \\ 

\begin{minipage}{\textwidth}
\begin{center}
\centering
\small
\begin{tabular}{rrrrrrrrr} 
 \\ \hline
Dec & E\_RA & E\_Dec & Total\_flux & E\_Total\_flux & E\_Total\_flux& Total\_flux & E\_Total\_flux & E\_Total\_flux\ \\ 
   &  &  & \_{Gaussian} & \_{Gaussian} & \_{Gaussian} & \_Source & \_Source & \_Source \\ 
      &  &  &  & \_PyBDSF &  &  & \_PyBDSF &  \\ 
 \degree & \arcsec & \arcsec & mJy & mJy & mJy & mJy & mJy & mJy \\ \hline \hline 
 10.302451 & 1.75 & 1.04 & 3.560  & 0.666 & 1.003 & 3.560 & 0.666 & 1.003 \\  
 10.327487 & 0.24 & 0.23 & 14.890& 0.562 & 1.642 & 14.890 & 0.562 & 1.642 \\  
 11.043854 & 3.45 & 4.50 & 7.158  & 1.613 & 1.898 & 7.158 & 1.613 & 1.898 \\  
 11.107845 & 1.10 & 1.26 & 3.163  & 0.628 & 0.957 & 3.163 & 0.628 & 0.957 \\  
 10.652017 & 1.67 & 2.52 & 2.563  & 0.695 & 0.972 & 2.563 & 0.695 & 0.972 \\  
 11.781003 & 1.71 & 1.32 & 4.822  & 0.889 & 1.222 & 4.822 & 0.889 & 1.222 \\  
 12.755545 & 2.17 & 1.26 & 13.814& 1.632 & 2.195 & 13.814 & 1.632 & 2.195 \\  
 13.699465 & 0.05 & 0.05 & 73.474& 0.572 & 5.672 & 73.474 & 0.572 & 5.672 \\  
 11.474149 & 0.11 & 0.11 & 41.977 & 0.760 & 3.521 & 41.977 & 0.760 & 3.521 \\  
 10.954526 & 2.36 & 2.15 & 2.343  & 0.752 & 1.003 & 2.343 & 0.752 & 1.003 \\   \hline 
  \end{tabular} 
 \end{center}
\end{minipage} \\ \\ 

\begin{minipage}{\textwidth}
\begin{center}
\centering
\small
\begin{tabular}{rrrrrrrrrrr} 
\\ \hline
Peak\_flux & E\_Peak\_flux & Maj & E\_Maj & Min & E\_Min & PA & E\_PA & DC\_Maj & E\_DC\_Maj & DC\_Min \\ 
 mJy/beam & mJy/beam & \arcsec & \arcsec & \arcsec & \arcsec & \degree & \degree & \arcsec & \arcsec & \arcsec \\ \hline \hline  
   2.968 & 0.336 & 32.24 & 4.23 & 23.31 & 2.27 & 105.09& 17.11 & 0.00 & 4.23 & 0.00 \\  
 14.881 & 0.324 & 25.44 & 0.56 & 24.64 & 0.53 & 80.74  & 29.91 & 0.00 & 0.56 & 0.00 \\  
 2.060   & 0.368 & 60.15 & 12.06 & 36.16 & 5.76 & 33.30& 20.82 & 54.70& 12.06 & 26.12 \\  
 3.318   & 0.374 & 25.58 & 3.03 & 23.33 & 2.50 & 158.81& 51.49 & 0.00 & 3.03 & 0.00 \\  
 1.971   & 0.334 & 31.86 & 5.95 & 25.56 & 3.92 &    0.26 & 35.41 & 19.73 & 5.95 & 5.13 \\  
 3.460   & 0.407 & 32.56 & 4.16 & 26.79 & 2.91 & 111.73 & 27.96 & 20.82 & 4.16 & 9.61 \\  
 5.772   & 0.495 & 53.55 & 5.57 & 27.97 & 1.98 & 116.00 & 8.56   & 47.34 & 5.57 & 12.54 \\  
 71.689 & 0.325 & 25.49 & 0.12 & 25.17 & 0.11 & 100.85 & 17.02 & 4.79 & 0.12 & 2.89 \\  
 40.863 & 0.431 & 25.47 & 0.27 & 25.25 & 0.26 & 154.41 & 43.48 & 4.85 & 0.27 & 3.27 \\  
 2.020   & 0.392 & 28.78 & 5.94 & 25.24 & 4.60 & 124.68 & 64.18 & 14.23 & 5.94 & 3.17 \\ \hline 
  \hline 
 \end{tabular} 
 \end{center}
\end{minipage} \\ \\ 

\begin{minipage}{\textwidth}
\begin{center}
\centering
\small
\begin{tabular}{rrrrrrrrrr} 
\\ \hline
E\_DC & DC\_PA & E\_DC & S\_Code & Separation & Noise & Gal\_lon & Gal\_lat \\ 
 &&\_PA  &  & \_Tile\_Centre &  &  &  \\ 
  \arcsec & \degree & \degree &  & \degree & mJy/beam & \degree & \degree \\ \hline \hline  
 2.27 & 0.00 & 17.11 & S & 3.8431 &   0.329 & 107.518268 & -51.404165 \\  
 0.53 & 0.00 & 29.91 & S & 3.8262 &   0.324 & 107.524147 & -51.379281 \\  
 5.76 &33.30 & 20.82 & S & 3.4484 &  0.367 & 107.793667 & -50.683253 \\  
 2.50 & 0.00 & 51.49 & S & 3.4105 &    0.379 & 107.801283 & -50.618566 \\  
 3.92 & 0.26 & 35.41 & S & 3.6220 &    0.319 & 107.609691 & -51.058232 \\  
 2.91 & 111.73 & 27.96 & S & 3.1691 & 0.385 & 108.040448 & -49.962954 \\  
 1.98 & 116.00 & 8.56 & S & 3.0707 &   0.488 & 108.394965 & -49.015996 \\  
 0.11 & 100.85 & 17.02 & S & 3.2624 & 0.323 & 108.725686 & -48.097792 \\  
 0.26 & 154.41 & 43.48 & S & 3.2441 & 0.428 & 107.899652 & -50.256826 \\  
 4.60 & 124.68 & 64.18 & S & 3.4531 & 0.380 & 107.696621 & -50.760190 \\   \hline 
  \end{tabular} 
\end{center} 
 \caption{{The first 10 lines from the final {Gaussian} component catalogue. The columns are described in Section~\ref{sec:gauscat}.}}
 \label{tab:compcat}
 \end{minipage}
\end{table*} 

\subsubsection{Source Catalogue}
\label{sec:srccat}

For the Source catalogue, we define the following columns:
\begin{itemize}
    \item \texttt{Source\_Name} - The name of the source given in the IAU convention JHHMMSS.S$\pm$DDMMSS with the prefix RACS-DR1\footnote{The DR1 has been added as we named this Data Release 1.}
    \item \texttt{Source\_ID} - The ID of the source given by the RACS tile ID added to the \texttt{Src\_ID} generated by \texttt{PyBDSF}
    \item \texttt{Tile\_ID} - The ID of the tile that the source was located in.
    \item \texttt{SBID} - The ID of the scheduling block associated with the observation.
    \item \texttt{Obs\_Start\_Time} - The time that the pointing observation started as Modified Julian Day (MJD) expressed in days.
    \item \texttt{N\_Gaus} - The number of Gaussian components that were used to fit the source
    \item \texttt{RA} and \texttt{Dec} (and errors) - The J2000 position of the source and its associated errors
    \item \texttt{Total\_flux\_Source} - The total flux density of the entire source (i.e. the sum of the Gaussian components and the \texttt{Total\_Flux} column in the \texttt{PyBDSF} source catalogue).
    \item \texttt{E\_Total\_flux\_Source\_PyBDSF} - The error on the total flux density from the \texttt{E\_Total\_Flux} column in \texttt{PyBDSF}.
     \item \texttt{E\_Total\_flux\_Source} - The combined error on the total flux density derived by summing in quadrature the error from PyBDSF with the errors of flux density from Equation 7 of \cite{RACS}.
    \item \texttt{Peak\_flux} (and error) - The modelled peak flux density for the source and its associated error from \texttt{PyBDSF}
    \item \texttt{Maj}, \texttt{Min} and \texttt{PA} (and errors) - The major axis, minor axis and position angle of the source fit by \texttt{PyBDSF}
     \item \texttt{DC\_Maj}, \texttt{DC\_Min} and \texttt{DC\_PA} (and errors) - The deconvolved major axis, minor axis and position angle of the source
      \item \texttt{S\_Code} - The code from \texttt{PyBDSF} which defines whether a source is a single (S), multiple (M) or complex (C) source. A single source (S) is a single Gaussian source corresponding to a single island. A multiple (M) is where a single source is comprised of multiple Gaussians. A complex source (C) is a source where there are multiple Gaussians which form multiple sources within an island.
      \item \texttt{Separation\_Tile\_Centre} - The distance between the source and the centre of the tile it is located in.
    \item \texttt{Noise} - The rms noise within the island boundary, quoted from the \texttt{Isl\_rms }column in \texttt{PyBDSF}.
    \item \texttt{Gal\_lon} and \texttt{Gal\_lat} - The Galactic longitude and latitude of the source in degrees
	\item \texttt{Flag\_Close} - {All sources where there was another source within 25\arcsec \ are flagged with a `C'. For 2 pairs of sources, these were so closely located that the \texttt{Source\_Name} was identical. This is only 2 \texttt{Source\_Name}'s out of $\sim$2 million and so we have flagged these with `CD' in this column. For Sources with no match within 25\arcsec \ have `-' in this column.}\footnote{These close sources that are not specified to be the same source by \texttt{PyBDSF} likely arise from \texttt{PyBDSF} fitting components during the atrous mode and not associating these with a co-located source. This affects $\sim$850 sources.}
	\end{itemize}
Unless specified, associated are as described in the \texttt{PyBDSF} documentation.

\subsubsection{{Gaussian} Component Catalogue}
\label{sec:gauscat}
For the {Gaussian} component catalogue, the associated columns are:
\begin{itemize}
    \item \texttt{{Gaussian\_ID}} -The ID corresponding to the Gaussian component constructed as the RACS tile ID added {to a unique Gaussian ID for the Gaussian components} in the individual tile
    \item {\texttt{Source\_ID}, \texttt{Tile\_ID}, \texttt{SBID}, \texttt{Obs\_Start\_Time} and \texttt{N\_Gaus} - as above, describing the source associated with this {Gaussian} component}
    \item \texttt{RA/Dec} (and errors) - The J2000 position of the Gaussian component and its associated errors
    \item \texttt{Total\_Flux\_{Gaussian}} (and errors) - The modelled total flux density of each individual {Gaussian} component and the associated errors (similar to as described above for the source but now for the component flux density).
    \item \texttt{Total\_Flux\_Source} (and errors) - Total flux densities and errors as described for the source catalogue
    \item \texttt{Peak\_Flux} (and error) - The modelled peak flux density of the Gaussian component and its associated error.
    \item \texttt{Maj/Min/PA} (and error) - The major and minor axes of the source (FWHM) and the position angle of the Gaussian component used to model the source
    \item \texttt{Maj\_DC/Min\_DC/PA\_DC} (and errors) - The deconvolved source sizes and position angle of the Gaussian component
    \item \texttt{S\_Code} - as in source catalogue
    \item \texttt{Separation\_Tile\_Centre} - The distance between the Gaussian component and the centre of the pointing it is located in
    \item \texttt{Noise} - as in source catalogue
    \item \texttt{Gal\_lon} and \texttt{Gal\_lat} - The Galactic longitude and latitude of the Gaussian component
\end{itemize}
More information on how the parameters in the source (*srl.fits) and Gaussian component (*gaul.fits) catalogues {are} produced by \texttt{PyBDSF} can be found through the \texttt{PyBDSF} documentation\footnote{\url{https://www.astron.nl/citt/pybdsf/write_catalogue.html\#definition-of-output-columns}.}.

{We note here that other work may use differing terminology to the source/Gaussian definitions used in this work. For example, ``source" in other work may refer to the final radio object where separated lobes and components that have not been identified by \texttt{PyBDSF} as differing sources but that actually come from the same physical object are combined together. This process of combining ``sources" (as defined here) into the same physical object often relies on a combination of visual identification and either machine learning methods or likelihood ratios \citep[see e.g.][]{Banfield2015, Williams2019, Galvin2020}. The process of combining sources into objects for RACS, however, is beyond the scope of this work.}

\section{Comparisons with Previous Radio Surveys}
\label{sec:previous_surveys}
Having completed the construction of a final catalogue, we now make comparisons with previous radio surveys at various radio frequencies in order to validate the values determined from RACS.

\subsection{Comparison Images}
We begin with a visual comparison for a handful of RACS sources and their counterpart regions in SUMSS, NVSS and TGSS-ADR, to indicate the difference in image resolution and baseline sensitivity. We include a comparison image from the IR wavelength AllWISE survey \citep{Cutri2013}, to make comparisons for a nearby resolved galaxy. As all these surveys have different sky coverage, there is only a narrow declination window ($\delta=-$40\degree \ to $-$30\degree) where it is possible to make a comparison with all four surveys. 
To obtain these images we make use of the cutout servers for each of the respective surveys\footnote{{ \url{https://www.cv.nrao.edu/nvss/postage.shtml} \newline \url{https://vo.astron.nl/tgssadr/q\_fits/cutout/form} \newline \url{https://irsa.ipac.caltech.edu/applications/wise/} \newline for SUMSS, images are available through SkyView \newline \url{https://skyview.gsfc.nasa.gov/}}}.

\begin{figure*}
    \centering
    \includegraphics[width=18cm]{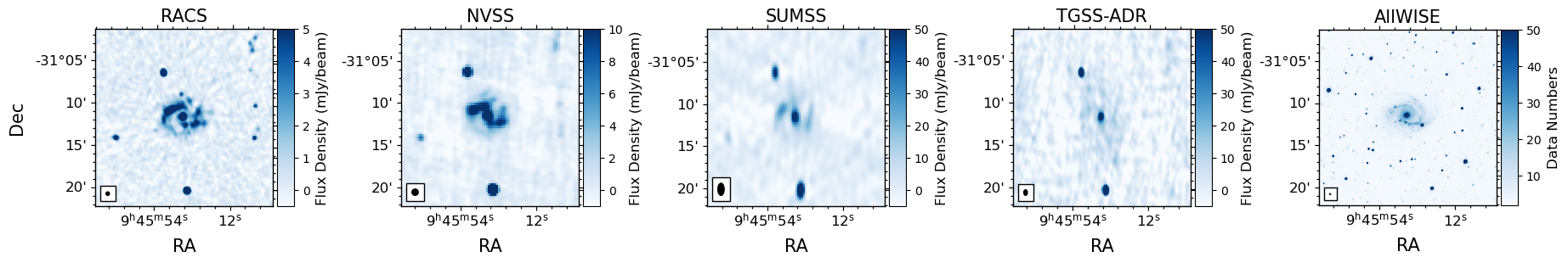} \\
    \includegraphics[width=18cm]{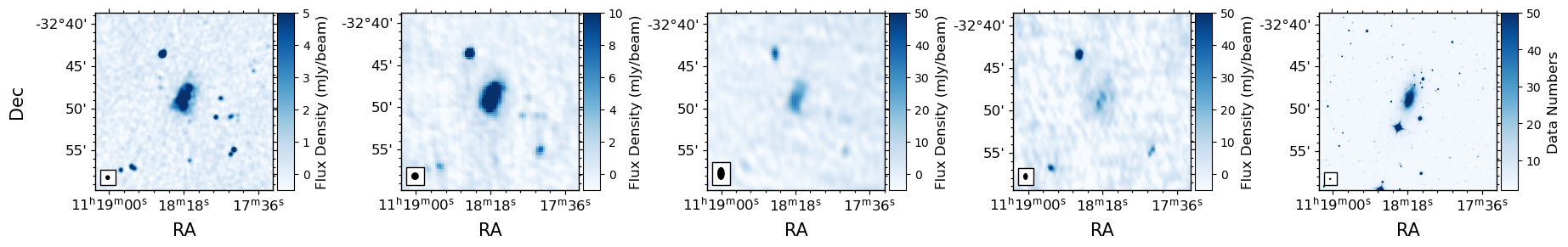}\\
    \includegraphics[width=18cm]{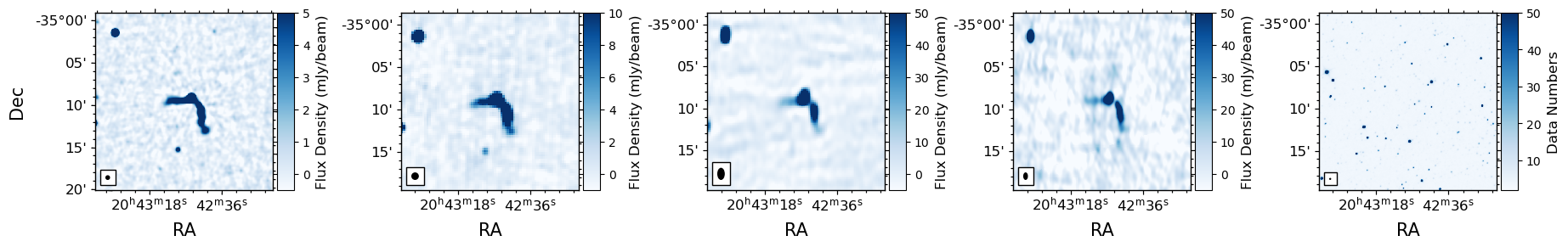}\\
    \includegraphics[width=18cm]{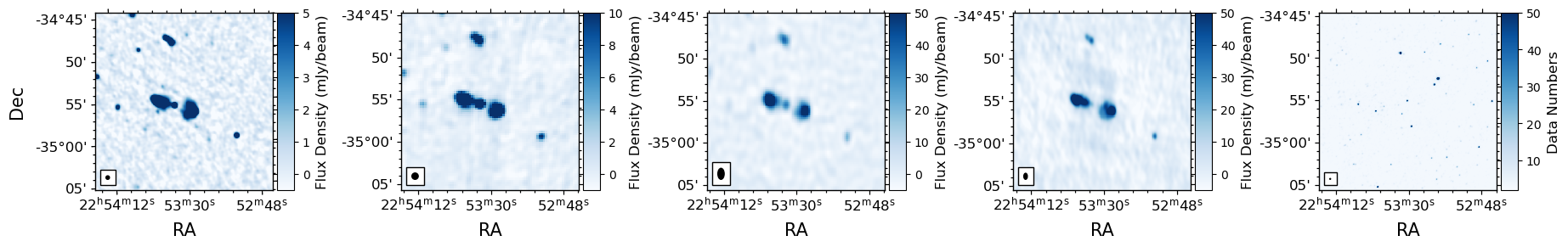}
    \caption{{Comparison images between RACS} at 25\arcsec \ (left), NVSS (centre left), SUMSS (centre), TGSS-ADR (centre right) and AllWISE W1 band (right) around four sources within the declination range $-$40\degree \ to $-$30\degree. The box in the bottom left of each panel indicates the PSF size for each observation. The flux density scales varies from image to image depending on its sensitivity.}
    \label{fig:im_comparisons}
\end{figure*}

Figure~\ref{fig:im_comparisons} demonstrates the higher resolution and increased sensitivity of RACS compared to SUMSS and NVSS. The sensitivity of ASKAP to extended emission is shown to be especially important (see the upper panel) to observe the structure in the spiral arms of the resolved galaxy NGC2997. These four cutouts highlight the improvement of RACS on previous large sky southern radio surveys. These images aim to indicate the quality that can be achieved with RACS. On the other hand, there may be regions, for example around bright sources, with poorer sensitivity compared to other surveys due to the snapshot nature of these observations and difficulties with the image processing.

Images in these regions will be improved with further observations of RACS as well as in the future with surveys such as the Evolutionary Map of the Universe {\citep[EMU;][]{Norris2011, EMU}}.

\subsection{Flux Offsets, Astrometric Offsets and Spectral Indices}
\label{sec:offsets}
It is important to ensure an accurate flux scale and accurate astrometry compared to previous observations as well as to investigate how the measured spectral index compares {to our knowledge of the radio source population}. We therefore compare our results to five previous large area radio surveys: GLEAM, NVSS, SUMSS and TGSS-ADR. Each of these surveys have different angular resolutions, operate at different frequencies and observe different (although often overlapping) regions of sky. Due to the differences in resolution and sensitivity, we restrict comparison to unresolved, high signal-to-noise, isolated sources. This ensures differences in the angular resolution, noise and sensitivity to extended emission do not affect our comparisons.

\subsubsection{Identifying Unresolved Sources}
\label{sec:unresolved}
\begin{figure*}
    \centering
    \includegraphics[width=18cm]{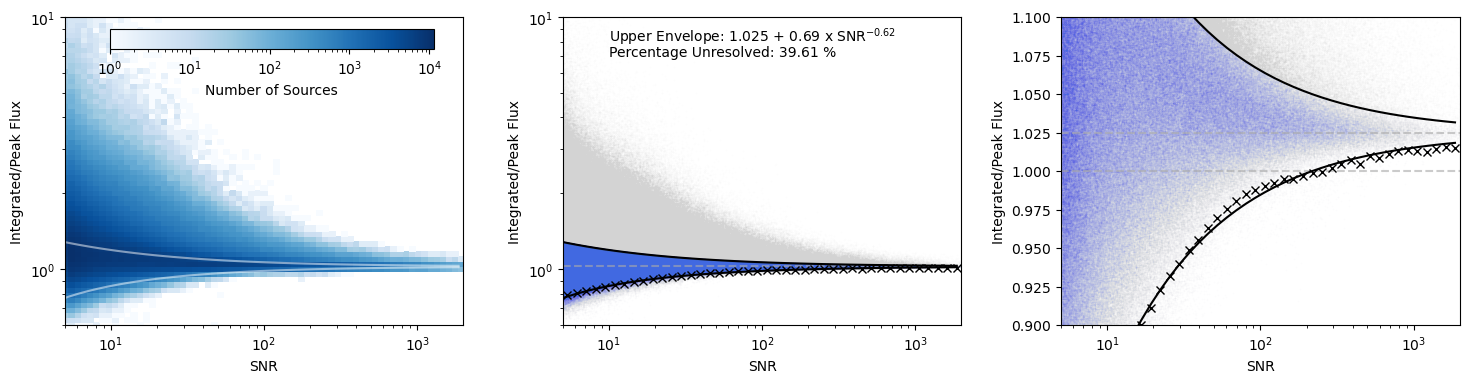}
    \caption{{The ratio of integrated to peak flux as a function of SNR for single component sources at $\geq5\sigma$ and the envelope (grey in left figure; black otherwise) used to define unresolved sources. Points in blue indicate those sources believed to be unresolved and grey indicates those sources believed to be resolved as defined by the envelope described in Section \protect \ref{sec:unresolved}. The black crosses indicates the $S_T/S_P$ values in which 95\% of sources below $S_T/S_P=1.025$ are included within the envelope. The grey dashed line indicates the ratio $S_T/S_P=1.025$. The right panel shows a closer view of the left panel around $S_T/S_P \sim 1$ (also given by a dashed line).}}
    \label{fig:envelope}
\end{figure*}

To select unresolved sources we follow {a previously-employed method \citep{Bondi2008, Smolcic2017, LoTSS} by defining}
an envelope to distinguish unresolved sources from those which are resolved. To construct this envelope, we used the Gaussian component catalogue and selected those components that were classified as single sources, and detected at SNR$\geq$5, where the SNR was defined as the peak flux of the Gaussian component divided by its island rms noise. We then considered how the ratio of the integrated flux density ($S_T$) to peak ($S_P$) flux density as a function of SNR; see Figure~\ref{fig:envelope}. 

The total flux-density $S_T$ of an unresolved source with peak brightness $S_P$= $S$ mJy/beam is $S_T$ = $S$ mJy, by construction. Therefore, if a source is unresolved and the synthesized beam size is a correct representation of the image resolution, the ratio of the integrated to peak flux ($S_T/S_P$) should be identically 1. This ratio, however, often has scatter around 1, especially at low SNR where faint sources are more affected by noise at the source position. For our data we find that as the SNR increases, $S_T/S_P$ tends to a value of 1.025, as illustrated in Figure~\ref{fig:envelope} (right panel). The source of the discrepant value of $S_T/S_P$ is unimportant for our analysis here, but must lie in some unmodelled source smearing due to effects such as uncorrected gain errors or astrometric mismatches between overlapping beams. Following the methods of \citep{Bondi2008, Smolcic2017, LoTSS} we expect values of $S_T/S_P$, as a function of SNR, to lie predominantly between the envelopes described by:

\begin{equation}
    \frac{S_T}{S_P}_{\pm} = 1.025 \pm A \times \textrm{SNR}^{-B}.
    \label{eq:envelope_eq}
\end{equation}

As resolved sources will have elevated values of $S_T/S_P$, we determine values for A, B from the lower envelope $S_T/{S_P}_{-}$ and declare sources with $S_T/S_P$ > $S_T/{S_P}_{+}$ to be resolved. To generate this fit, we use equally spaced logarithmic bins in SNR. For each bin with 100 {sources} or more, we find the $S_T/S_P$ ratio that contains 95\% of the sources with $S_T/S_P<1.025$, indicated by the black crosses on Figure~\ref{fig:envelope}. These points are fit to Equation~\ref{eq:envelope_eq} using the \texttt{Scipy} function \texttt{curve\_fit}. This fit to the lower envelope is determined to be: 1.025 $-$ 0.69 $\times$ SNR$^{-0.62}$. We reflect this envelope about $S_T/S_P$ = 1.025 and define the upper envelope:  $S_T/S_P$ = 1.025 + 0.69 $\times$ SNR$^{-0.62}$. {Sources below the upper and lower envelopes are determined to be unresolved. Unresolved components are shown as blue points in Figure~\ref{fig:envelope} and resolved sources in grey. From this we estimate approximately 40\% of {RACS} sources are unresolved {{at 25\arcsec} resolution}, and should therefore also be unresolved in the comparison catalogs.}

\subsubsection{Matching catalogues}
\label{sec:matching_cats}
For comparison with other catalogues, RACS sources are selected according to the following criteria:

\begin{enumerate}
    \item Are isolated within an angular separation of $N_{ISO}$\arcsec. The value of $N_{ISO}$ is given as twice the poorer resolution (using the FWHM) of the two catalogues being compared. We apply the same `isolated' criterion for the comparison radio survey.
    \item Have a peak SNR in RACS $\geq10$ 
    \item Satisfy the unresolved envelope criterion as described above.
    \item Match the comparison radio catalogue within an angular separation of $N_{match}$\arcsec. Here $N_{match}$ is taken to be 10\arcsec. This value corresponds to 4 pixels in the RACS images and allows for variation in the positions measured of sources, given NVSS and SUMSS have an angular resolution $\sim$2 times poorer {than} RACS.
\end{enumerate}

The resolution and frequency for each of the surveys we compare {to RACS} are shown in Table~\ref{tab:offsets}. We use sources which satisfy the match criteria to consider the offsets in flux and astrometry, as well as the measured spectral indices. The spectral index ($\alpha$) is used to define the broadband radio emission as a power law of the form $S_{\nu} \propto \nu ^ {\alpha}$, where $S_{\nu}$ is the flux density at a frequency, $\nu$. Typically $\alpha$ is found in catalogues to have an average value of $-$0.7 to $-$0.8 in the synchrotron dominated regime \citep[see e.g.][]{Condon1992, SUMSS, Smolcic2017}.

\begin{table*}[]
    \centering
    \small
    \begin{tabular}{l  l  l l  l  l l}
        & AT20G  & GLEAM & NVSS & SUMSS & TGSS-ADR & TGSS-ADR-R \\ \hline \hline
       Frequency (MHz) & 20,000 & 200 (wide band) & 1,400 & 843 & 150 & 150 \\ 
       Resolution (arcsec)  & $\sim$10 & 120 & 45 & $\frac{45}{|\textrm{sin}\delta|}$ & 25 ($\delta \geq19^{\circ}$) & 25 ($\delta \geq19^{\circ}$)  \\ 
        &   &   &   &   & $\frac{25}{\textrm{cos}(\delta - 19^{\circ})}$ ($\delta <19^{\circ}$)  & $\frac{25}{\textrm{cos}(\delta - 19^{\circ})}$ ($\delta <19^{\circ}$)  \\ 
       Assumed 5$\sigma$ Limit (mJy) & - & 40 & 2.5 & - & 20 & 20 \\ 
        at observed frequency &  & & & & & \\ \\
       N Matches (Flux)  & - &  - & -  & \flsumssN & - & - \\ 
       Flux Ratio  &  - & - & - & $\flsumss^{+\flsumssperr}_{-\flsumssnerr}$ & - \\  \\
       N Matches (Astrometry)  & \raatN &  - & \ranvssN  & \rasumssN & - & -  \\ 
       RA Offset (\arcsec)  & $\raat^{+\raatperr}_{-\raatnerr}$  &  & $\ranvss^{+\ranvssperr}_{-\ranvssnerr}$ & $\rasumss^{+\rasumssperr}_{-\rasumssnerr}$ & & \\ 
       Dec Offset (\arcsec)   & $\decat^{+\decatperr}_{-\decatnerr}$ &  & $\decnvss^{+\decnvssperr}_{-\decnvssnerr}$ & $\decsumss^{+\decsumssperr}_{-\decsumssnerr}$ & & \\ \\
       N Matches & - &  \algleamN & \alnvssN  & - & \altgssN & \altgssRN \\
       (Alpha  - No Flux Cut) & & & & & & \\
       Alpha (No Flux Cut) & - & $\algleam^{+\algleamperr}_{-\algleamnerr}$ & $\alnvss^{+\alnvssperr}_{-\alnvssnerr}$ & -& $\altgss^{+\altgssperr}_{-\altgssnerr}$ & $\altgssR^{+\altgssRperr}_{-\altgssRnerr}$ \\ \\
       N Matches & - &  \alfgleamN & \alfnvssN  & - & \alftgssN & \alftgssRN \\ 
        (Alpha  - Flux Cut) & & & & & & \\
       Alpha (Flux Cut) & - & $\alfgleam^{+\alfgleamperr}_{-\alfgleamnerr}$ & $\alfnvss^{+\alfnvssperr}_{-\alfnvssnerr}$ & -& $\alftgss^{+\alftgssperr}_{-\alftgssnerr}$ & $\alftgssR^{+\alftgssRperr}_{-\alftgssRnerr}$ \\ \hline

    \end{tabular}
    \caption{{Measured flux density and astrometric offsets, as well as spectral index comparisons between RACS and GLEAM, NVSS, SUMSS and TGSS-ADR as also the rescaled TGSS-ADR \citep{TGSSR}. Offsets are quoted as the median value as well as the associated errors using the 16$^{\rm th}$ and 84$^{\rm th}$ percentiles.}}
    \label{tab:offsets}
\end{table*}

\subsubsection{Flux Offsets}
\label{sec:floffsets}
We make flux density comparisons using SUMSS due to its close proximity in frequency to RACS (843 MHz for SUMSS compared to 887.5 MHz for RACS). This minimizes any effect of spectral index {uncertainty} on flux density comparisons. For example, assuming a nominal spectral index of $\alpha=-0.8\pm0.1$ we expect the frequency differences between RACS and SUMSS to result in a flux offset of $\pm0.5$\%, increasing to $\pm$5\% at the frequency of NVSS resulting from the error in spectral index.

\begin{figure*}[h!]
\begin{center}
\includegraphics[width=8cm]{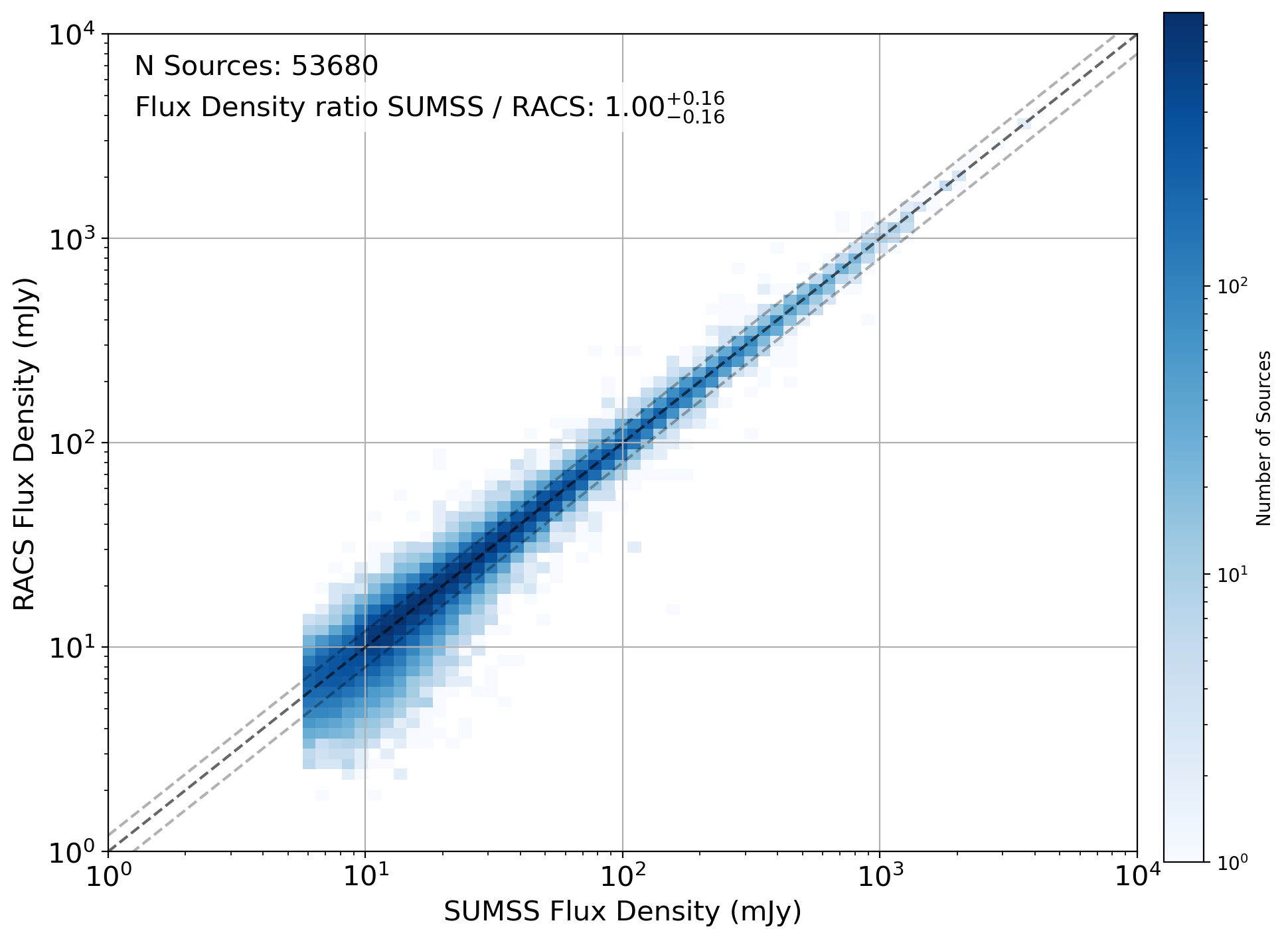}
\caption{{Flux density comparison between RACS and SUMSS at a frequency of 887.5 MHz (assuming $\alpha=-0.8$), for sources matched using the criteria described in Section~\ref{sec:matching_cats}. The black dashed line indicates a 1-to-1 relation, whilst the grey dashed lines indicate flux ratios of 80 or 120\%. }} 
\label{fig:flux_comparisons}
\end{center}
\end{figure*}

\begin{figure*}[h!]
\begin{subfigure}{0.5\textwidth}
\includegraphics[width=\textwidth]{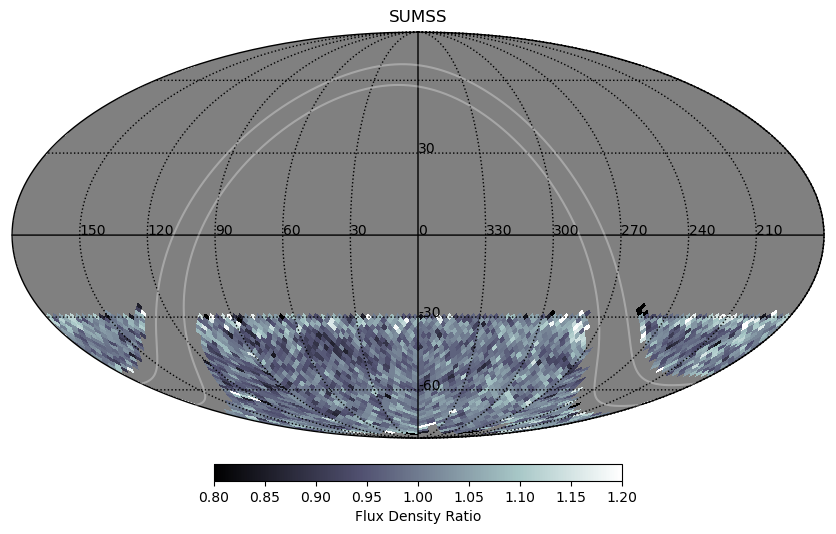}
\subcaption{SUMSS}
\end{subfigure}
\begin{subfigure}{0.5\textwidth}
\includegraphics[width=\textwidth]{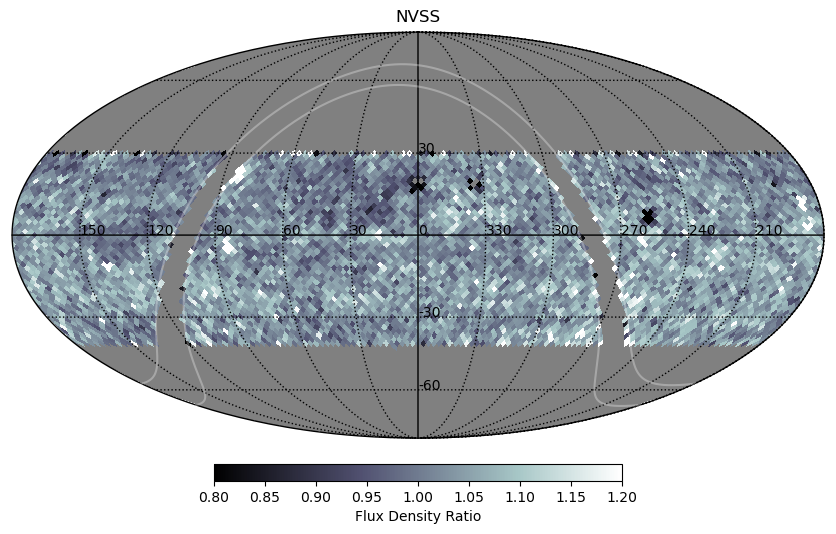}
\subcaption{NVSS}
\end{subfigure}
\caption{Flux density ratio comparison as a function of sky position between RACS and SUMSS (left) at a frequency of 887.5 MHz (assuming $\alpha=-0.8$), for sources as matched per the criteria in Section~\ref{sec:matching_cats}. We also include the comparison with NVSS (right) to allow for a comparison of the flux density ratio over the full sky. } \label{fig:flux_sumss}
\end{figure*}

Using the matching criteria described above, \flsumssN matched sources were identified. The comparison of total flux densities assuming a spectral index of $\alpha=-0.8$ can be seen in Figure~\ref{fig:flux_comparisons}. From this we find a median flux ratio of $\flsumss^{+\flsumssperr}_{-\flsumssnerr}$. The associated errors are quoted from the 16$^{\textrm{th}}$ and 84$^{\textrm{th}}$ percentiles. We therefore conclude that we have an accurate flux scale for our observations. This flux comparison as a function of position can also be seen in Figure~\ref{fig:flux_sumss}. We present this for both comparisons with SUMSS (Figure~\ref{fig:flux_sumss}, left) but also show this comparison with NVSS (Figure~\ref{fig:flux_sumss}, right). Whilst the difference in frequency compared to NVSS is larger, the two figures in Figure~\ref{fig:flux_sumss} combined show the flux offsets across the majority of the coverage of RACS. Figure~\ref{fig:flux_sumss} does not appear to show significant systematic variation in the ratios of flux density as a function of position.

\subsubsection{Astrometric Offsets}

\begin{figure*}[h!]
\begin{subfigure}{0.5\textwidth}
\includegraphics[width=0.9\textwidth]{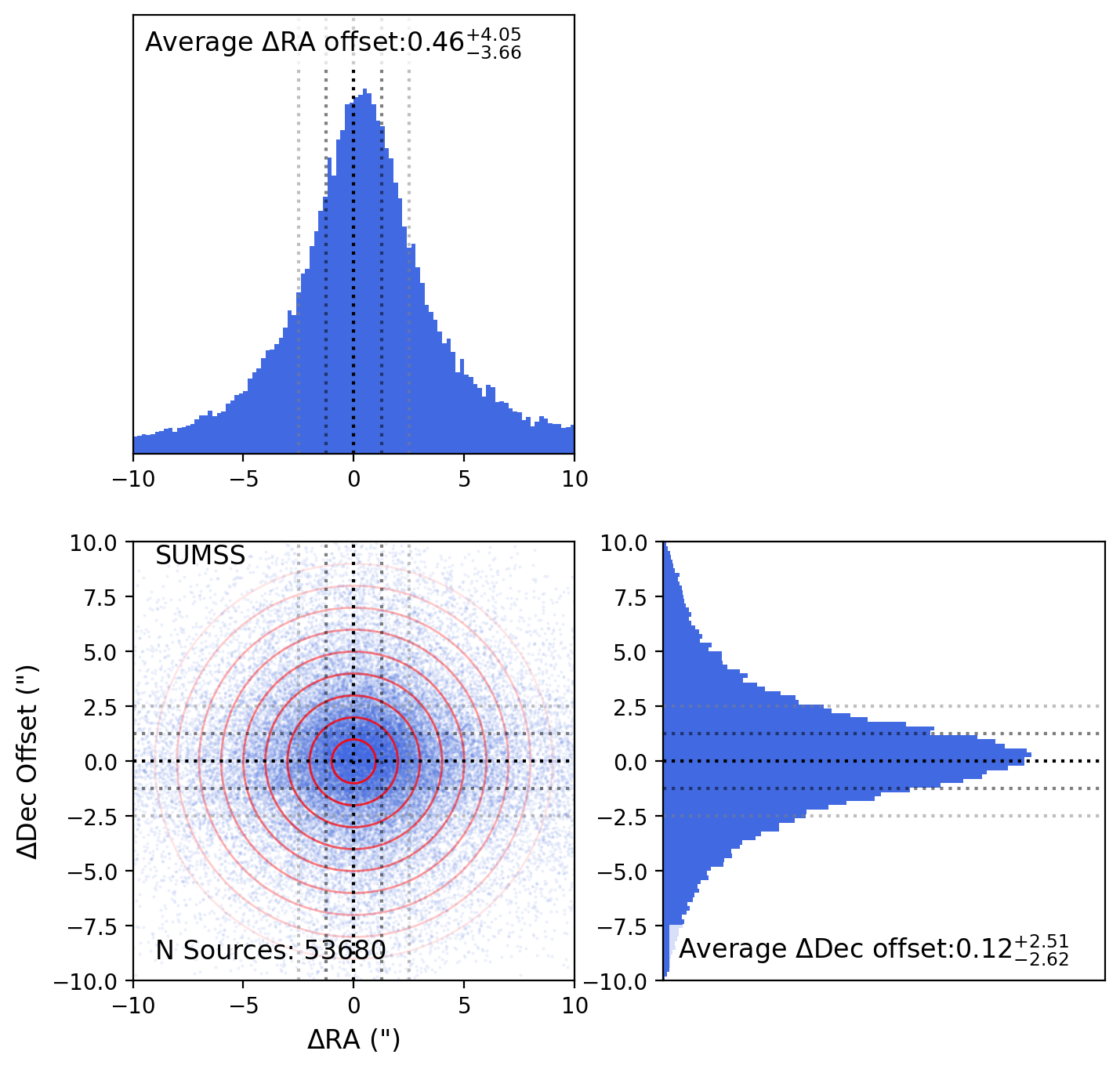}
\subcaption{SUMSS}
\end{subfigure}
\begin{subfigure}{0.5\textwidth}
\includegraphics[width=0.9\textwidth]{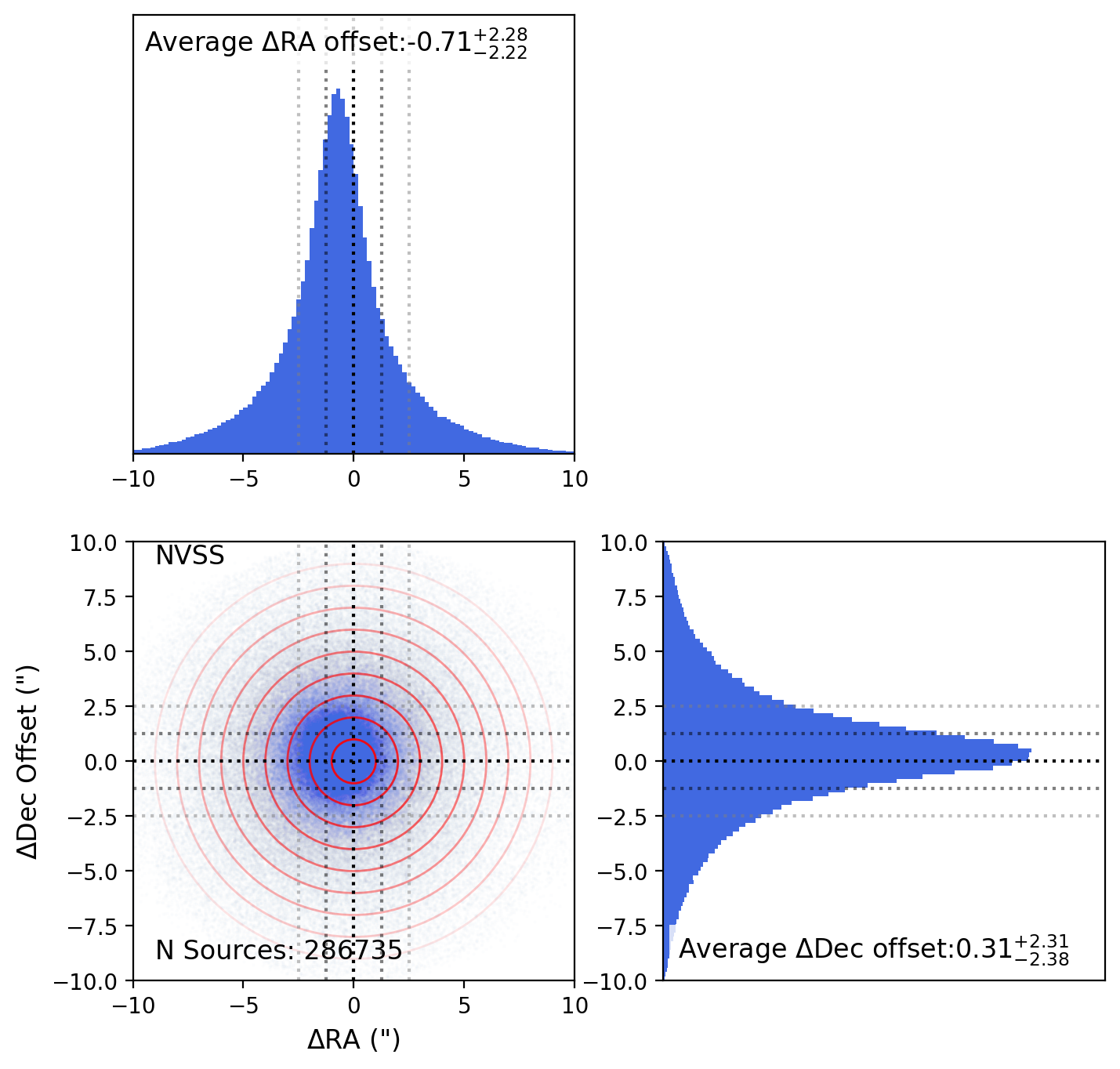}
\subcaption{NVSS}
\end{subfigure}
\begin{center}
\begin{subfigure}{0.5\textwidth}
\includegraphics[width=0.9\textwidth]{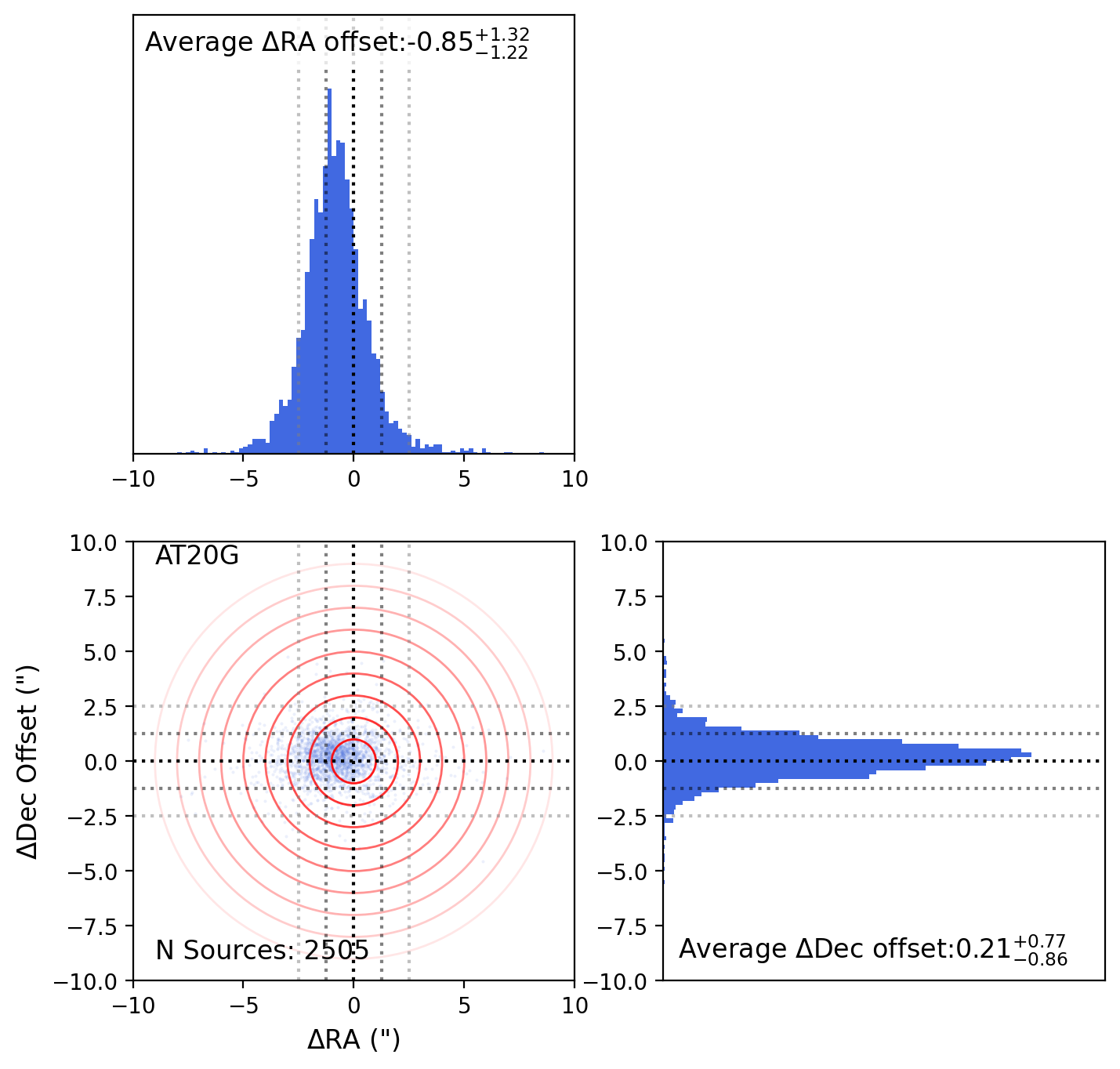}
\subcaption{AT20G}
\end{subfigure}
\end{center}
\caption{Comparison of the astrometric offsets between RACS and SUMSS (left), NVSS (centre) and AT20G (right) for sources matched with the criteria in Section~\ref{sec:matching_cats}. The red circles correspond to radii at 1\arcsec \ intervals from 1 to 9\arcsec \ and the grey dashed lines indicate the limits of $\pm$0.5 and $\pm$1 $\times$ the RACS pixel size. The black dashed lines indicate no astrometric offsets between the comparison surveys. } 
\label{fig:astrometry}
\end{figure*}

We assess the astrometry of RACS, using matches that satisfy the selection criteria described in Section~\ref{sec:matching_cats} for some of the catalogues described in Table~\ref{tab:offsets}. We define the RA offset to be: $\Delta \textrm{RA}$ = RA$_{\textrm{RACS}}$- RA$_{\textrm{Comp}}$ where ``Comp'' refers to the comparison survey. The Declination offset is defined in the same way. These astrometric offsets can be seen in Figure~\ref{fig:astrometry}. We compare to SUMSS and NVSS, but not to GLEAM due to its much larger PSF ($\sim$2\arcmin), nor to TGSS-ADR as it was tied to the astrometry of NVSS ($\delta\geq-35$\degree) and MGPS or SUMSS ($\delta\leq-35$\degree) to avoid residual astrometric errors from ionospheric interference at low frequencies. We also include a comparison to AT20G which, although it has far fewer comparison sources than NVSS and SUMSS,  provides a comparison with surveys at much higher frequencies.

From this we find small median systematic offsets in both RA and Dec, |Offset|$\lesssim$ 0.8\arcsec, where the RA value of RACS is systematically lower than NVSS and AT20G but larger than SUMSS. Here we find RA offsets (in \arcsec) of: $\raat^{+\raatperr}_{-\raatnerr}$ (AT20G), $\ranvss^{+\ranvssperr}_{-\ranvssnerr}$ (NVSS) and $\rasumss^{+\rasumssperr}_{-\rasumssnerr}$ (SUMSS). The Dec offset is smaller in magnitude than for RA. The measured Dec offsets (in \arcsec) are: $\decat^{+\decatperr}_{-\decatnerr}$ (AT20G), $\decnvss^{+\decnvssperr}_{-\decnvssnerr}$ (NVSS) and $\decsumss^{+\decsumssperr}_{-\decsumssnerr}$ (SUMSS). However as the pixel size of the images is 2.5\arcsec, these offsets are typically constrained within a pixel or two. Further discussion of the beam to beam accuracy in astrometry within the individual beam images can be found in Paper I.

\begin{figure*}[h!]
\begin{subfigure}{0.5\textwidth}
\includegraphics[width=8cm]{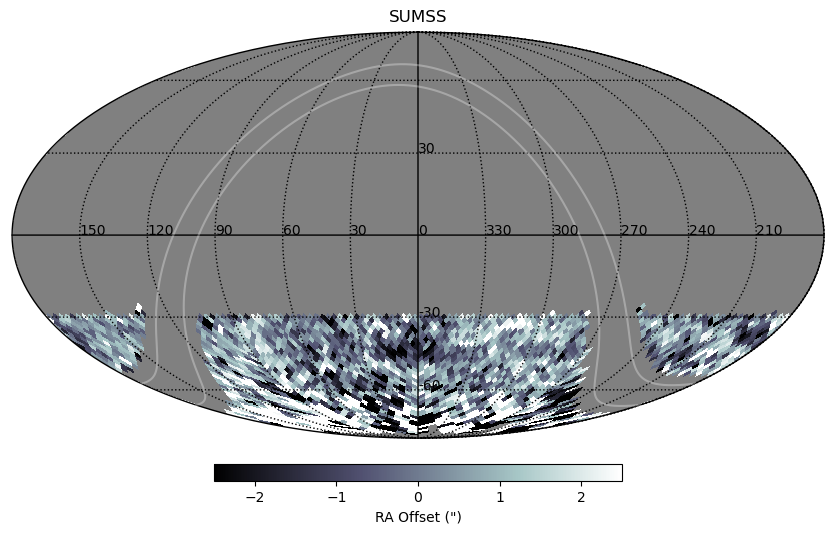}
\subcaption{SUMSS}
\end{subfigure}
\begin{subfigure}{0.5\textwidth}
\includegraphics[width=8cm]{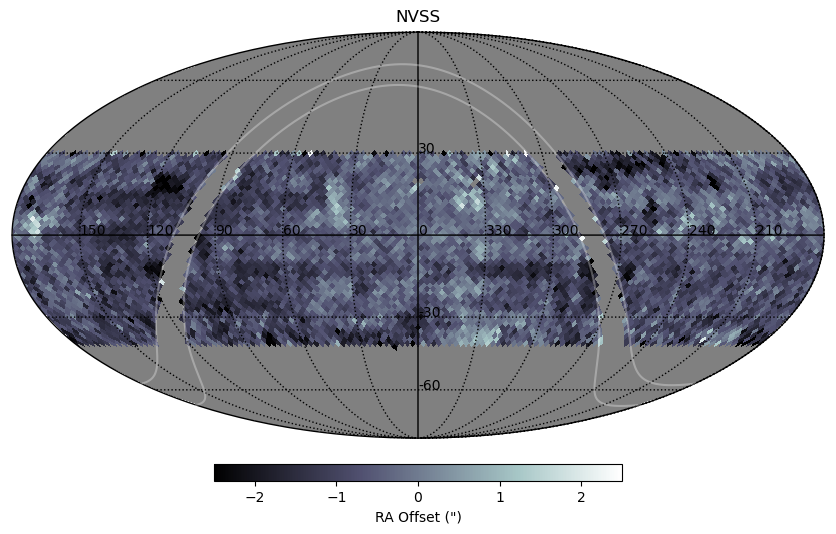}
\subcaption{NVSS}
\end{subfigure}
\caption{Comparison of the median RA offsets, for each \texttt{HEALPix} bin, between RACS and SUMSS (left) and NVSS (right) for sources matched with the criteria in Section~\ref{sec:matching_cats} as a function of position across the sky. } \label{fig:astrometry_RA}
\end{figure*}

\begin{figure*}[h!]
\begin{subfigure}{0.5\textwidth}
\includegraphics[width=8cm]{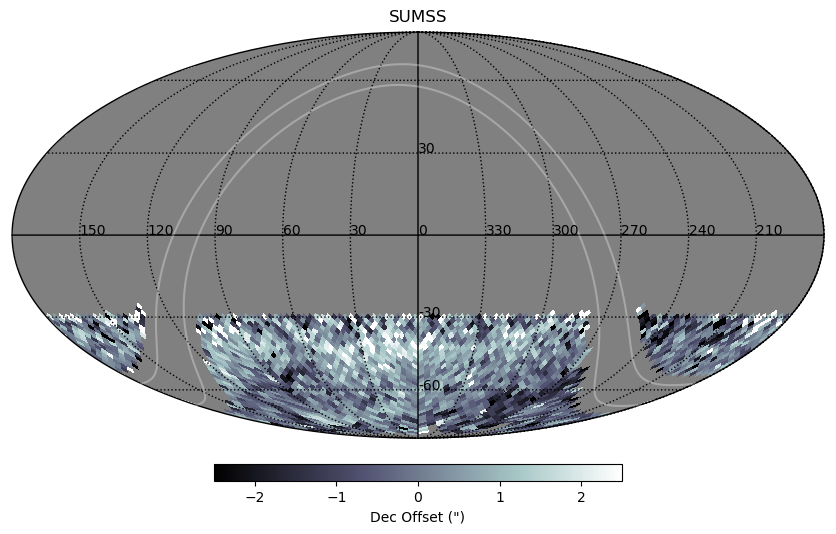}
\subcaption{SUMSS}
\end{subfigure}
\begin{subfigure}{0.5\textwidth}
\includegraphics[width=8cm]{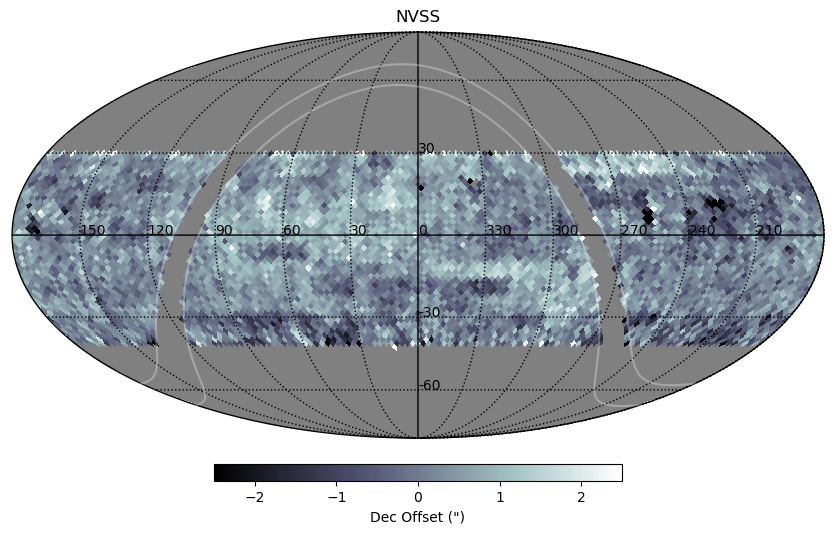}
\subcaption{NVSS}
\end{subfigure}
\caption{Comparison of the median Dec offsets, for each \texttt{HEALPix} bin, between RACS and SUMSS (left) and NVSS (right) for sources matched with the criteria in Section~\ref{sec:matching_cats} as a function of position across the sky.} \label{fig:astrometry_DEC}
\end{figure*}

The variation of astrometric offset with sky position can be seen in Figures~\ref{fig:astrometry_RA} and~\ref{fig:astrometry_DEC} for Right Ascension and Declination respectively.

\subsubsection{Spectral Index Comparisons}
\label{sec:specindex}
Finally, we compare the spectral index between RACS and radio surveys at other frequencies, assuming a power law spectral energy distribution (SED) as discussed in Section~\ref{sec:matching_cats}. We define $\alpha$ here as:

\begin{equation}
    \alpha^{\textrm{RACS}}_{\textrm{Comp}} = \frac{\log\left(\frac{S_{\textrm{RACS}}}{S_{\textrm{Comp}}}\right)}{\log\left(\frac{\nu_{\textrm{RACS}}}{\nu_{\textrm{Comp}}}\right)}.
\end{equation}

\begin{figure*}[h!]
\begin{subfigure}{0.5\textwidth}
\includegraphics[width=8.5cm]{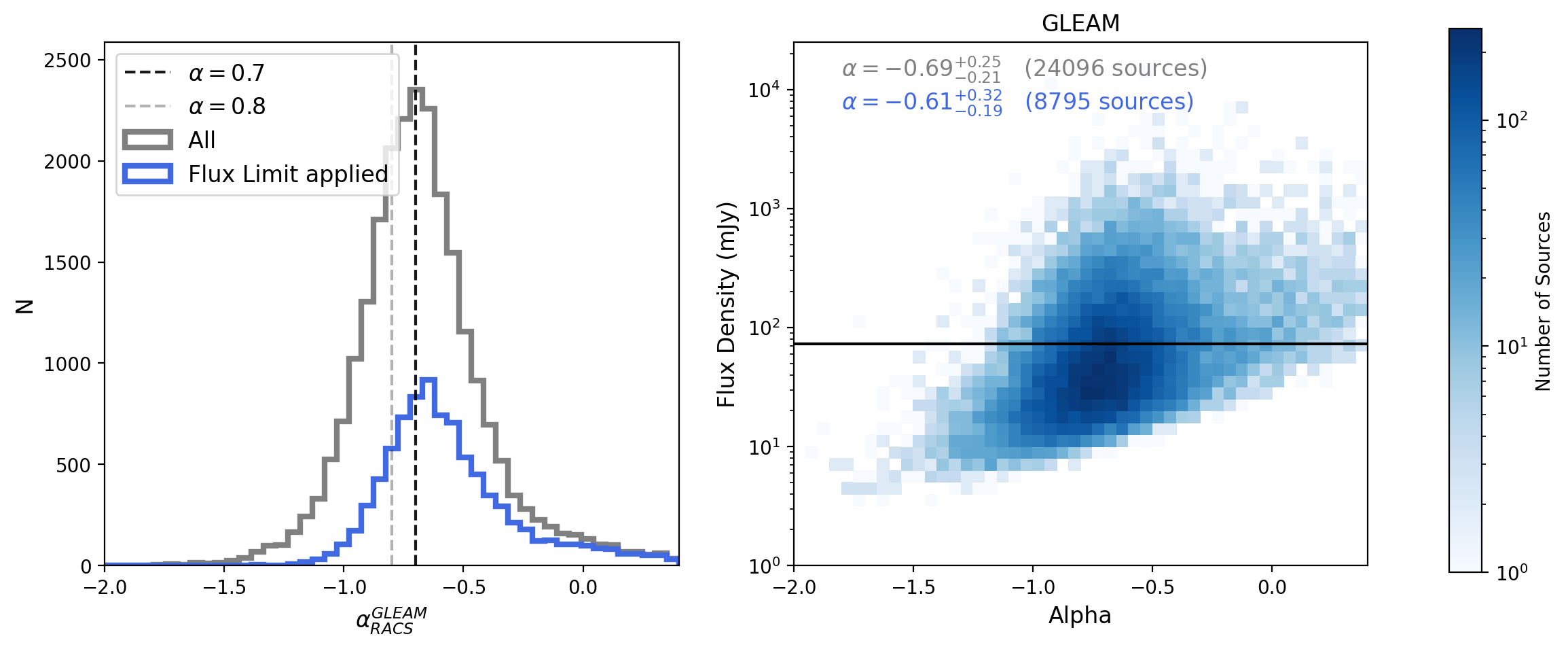}
\subcaption{GLEAM}
\end{subfigure}
\begin{subfigure}{0.5\textwidth}
\includegraphics[width=8.5cm]{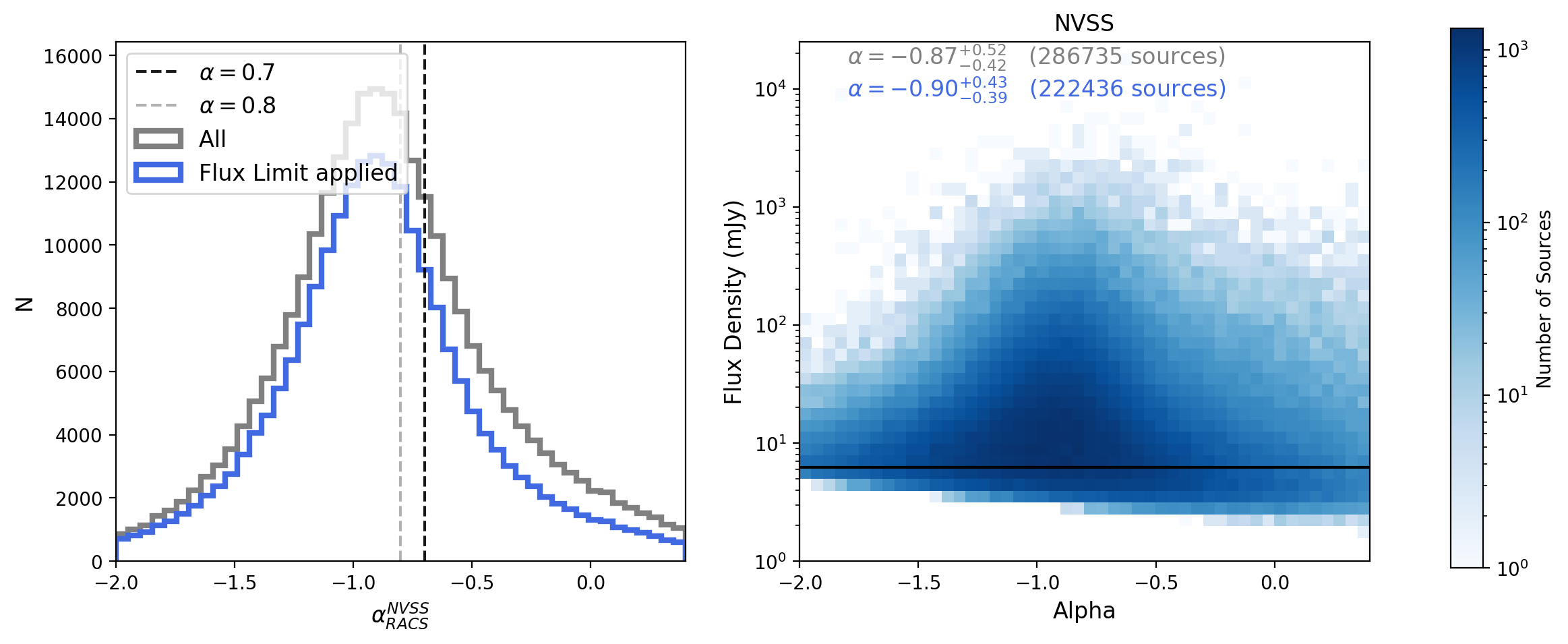}
\subcaption{NVSS}
\end{subfigure}
\begin{subfigure}{0.5\textwidth}
\includegraphics[width=8.5cm]{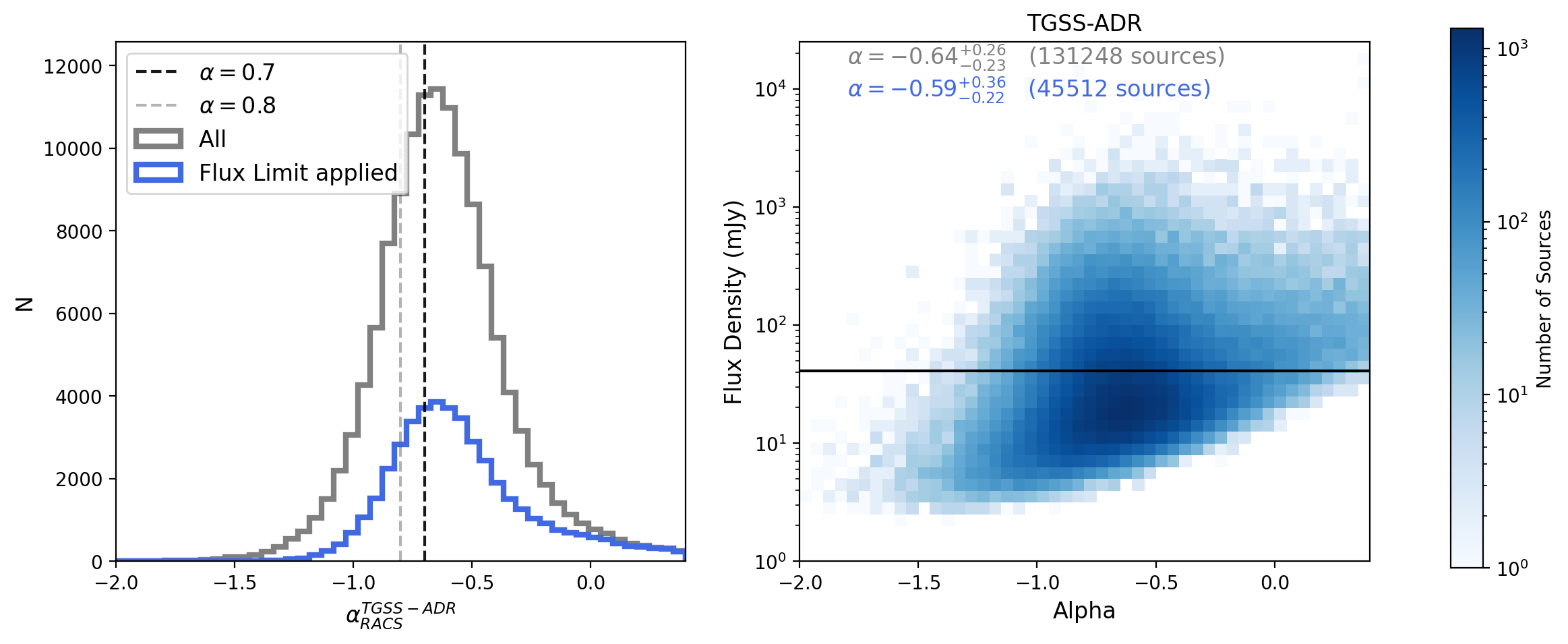}
\subcaption{TGSS-ADR}
\end{subfigure}
\begin{subfigure}{0.5\textwidth}
\includegraphics[width=8.5cm]{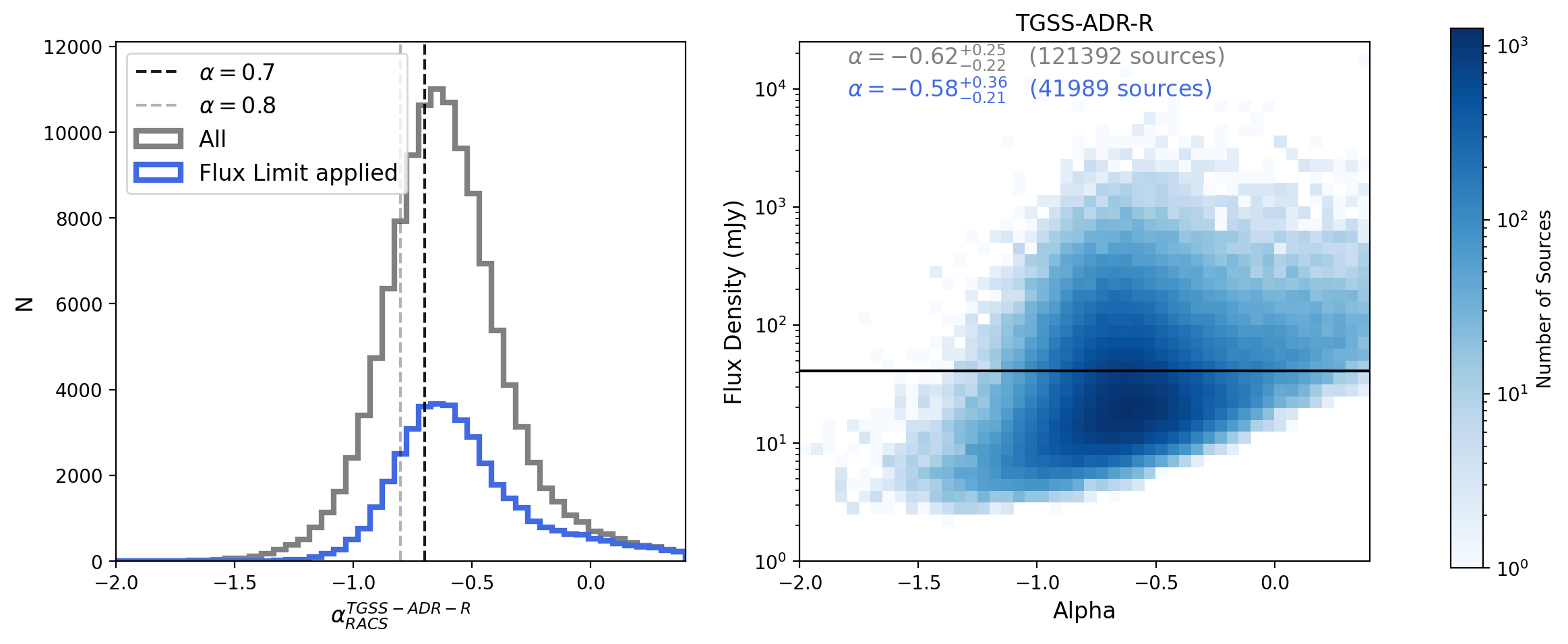}
\subcaption{TGSS-ADR-R}
\end{subfigure}
\caption{{Comparison of the spectral indices between RACS and (a) GLEAM, (b) NVSS, (c) TGSS-ADR and (d) the Rescaled TGSS-ADR catalogue. For each panel, the left panel shows the histogram distribution of $\alpha$, whilst the right panel shows the distribution of $\alpha$ with flux density. This is shown with (blue) and without (grey) a flux cut (see Section~\ref{sec:specindex}). A black solid line indicates where the flux density cut is applied. }} \label{fig:alpha}
\end{figure*}

It is important when measuring the spectral indices between matched catalogues that the sensitivity limits are considered. This will bias spectral indices to either lower or higher values depending on the sensitivity limits and frequencies of the comparison surveys. Therefore we consider the spectral index for our matched sources both with and without a flux density cut applied. To determine the flux density cuts to apply we assume the sensitivity limits of each survey to be the  5$\sigma$ sensitivity limits in Table~\ref{tab:offsets} or the approximate 10$\sigma$ sensitivity of RACS (taken here as 3 mJy). Using these sensitivity limits, we determine the flux density cuts that are necessary to ensure there is no bias within the range $\alpha = -0.8\pm 1.2$, which should encompass the majority of $\alpha$ values observed \citep[see e.g.][]{Smolcic2017,Tiwari2019}. We then apply any necessary flux cuts to avoid any bias in $\alpha$. This flux density cut greatly reduces the number of sources available for comparisons. The histogram distribution of these spectral indices can be seen in Figure~\ref{fig:alpha} (left) as well as the comparison of spectral index with flux density (right). This latter plot indicates the necessity of applying a flux limit cut when investigating the spectral index. 

For spectral index comparisons, we do not consider $\alpha^{\rm RACS}_{\rm SUMSS}$ due to the small frequency offset. However we add in a comparison to the rescaled TGSS-ADR catalogue from \cite{TGSSR}. This adjusted the flux scale of TGSS-ADR based on measurements from the GLEAM survey. For comparisons to this survey we will use the label `TGSS-ADR-R'.

From Figure~\ref{fig:alpha} we find a typical median $\alpha$ in the range $\sim-$0.6 to $-$0.9, encompassing the typical values expected of $\sim-$0.7 to $-$0.8. Without a flux cut applied, the median $\alpha$ and errors from the 16$^{\rm th}$ and 84$^{\rm th}$ percentiles are measured as: $\algleam^{+\algleamperr}_{-\algleamnerr}$ (GLEAM), $\alnvss^{+\alnvssperr}_{-\alnvssnerr}$ (NVSS), $\altgss^{+\altgssperr}_{-\altgssnerr}$ (TGSS-ADR) and $\altgssR^{+\altgssRperr}_{-\altgssRnerr}$ (TGSS-ADR-R). When a flux cut is applied, these are now measured as: $\alfgleam^{+\alfgleamperr}_{-\alfgleamnerr}$ (GLEAM), $\alfnvss^{+\alfnvssperr}_{-\alfnvssnerr}$ (NVSS), $\alftgss^{+\alftgssperr}_{-\alftgssnerr}$ (TGSS-ADR) and $\alftgssR^{+\alftgssRperr}_{-\alftgssRnerr}$ (TGSS-ADR-R). The comparisons with the low frequency surveys of TGSS-ADR and GLEAM are closer to $-$0.6 to $-$0.7 whilst the higher frequency comparison with NVSS is more similar to $-$0.9. This may suggest that the RACS fluxes are slightly larger than expected from previous surveys, however as shown in Section~\ref{sec:floffsets} we have a good flux comparison with SUMSS. \\ 

\noindent In general, these comparisons have shown that we have good systematic astrometric and flux characteristics compared to other surveys. Measurements of the spectral indices of RACS sources will also be improved with future RACS observations, which are planned for different frequency bands (see Paper I).

\section{Completeness}
\label{sec:comp_rel}
We consider the completeness of our catalogue as a function of flux density. It should be close to unity at high flux densities and will decline towards zero close to the detection threshold of the survey. Completeness is affected by both the variation of rms across the survey area, which affects the detection threshold, as well as the source finder itself. We therefore need to consider the completeness of this catalogue as a function of flux density. 

We consider the survey completeness in two forms, for unresolved sources and for a combination of both unresolved (point) and resolved sources. To investigate both of these, we use simulations in which we inject sources into the residual images (Image - Model from \texttt{PyBDSF}) and investigate the recovery of the injected sources with \texttt{PyBDSF}. These simulations are described below.

\subsection{Point Source Detection}
\label{sec:pointcomp}
First, to investigate the point source completeness, we injected Gaussians with the resolution of the images i.e. a circular PSF of 25\arcsec \ FWHM into our residual images. To consider the detection of sources at a variety of realistic radio flux densities, we use the simulated catalogues from SKADS \citep{Wilman2008, Wilman2010}. These simulations were created in preparation for the Square Kilometre Array (SKA) to provide realistic mock catalogues that reflect both observations from existing radio surveys as well as expectations of radio sources below current sensitivity limits. 

For 5 million random positions across the range $\delta = -$85\degree \ to +30\degree \ we find the closest tile for each random source. For each tile we consider the random sources which are closest to that tile and for each source we randomly choose a flux density from SKADS and scale this from 1400 MHz to 887.5 MHz, assuming a spectral index of $\alpha=-0.8$. We inject a Gaussian component with the simulated total flux density\footnote{ensuring the scaled SKADS flux density$\geq$ 0.5 mJy} into the residual image at the random position generated\footnote{{Due to this, a small fraction fewer than the 5 million source will be injected in practice, as some sources may be, for example, at locations where no image is available.}}. \texttt{PyBDSF} is then used to investigate the detection of simulated sources within the image, using the same parameters as in Section~\ref{sec:cat}. From this the comparison of detected sources across the image can be calculated. We repeat this for each tile within the observation. We repeat this method {10 times} to make multiple realisations of the simulated distribution of sources. We estimate the average completeness from the mean completeness in each flux bin considered and the error from the standard deviation across the 10 realisations.

Once all the output \texttt{PyBDSF} catalogues have been calculated for each field and for each simulation realisation, we compare the input sources to those measured. For each field, we match the input catalogue to the recovered catalogue and class those sources as ``recovered" as those output source which match to an input source within half the FWHM resolution of our images (i.e. 12.5 \arcsec). We then calculate the detection fraction in two methods.

\begin{figure*}[h!]
\begin{center}
\includegraphics[width=16cm]{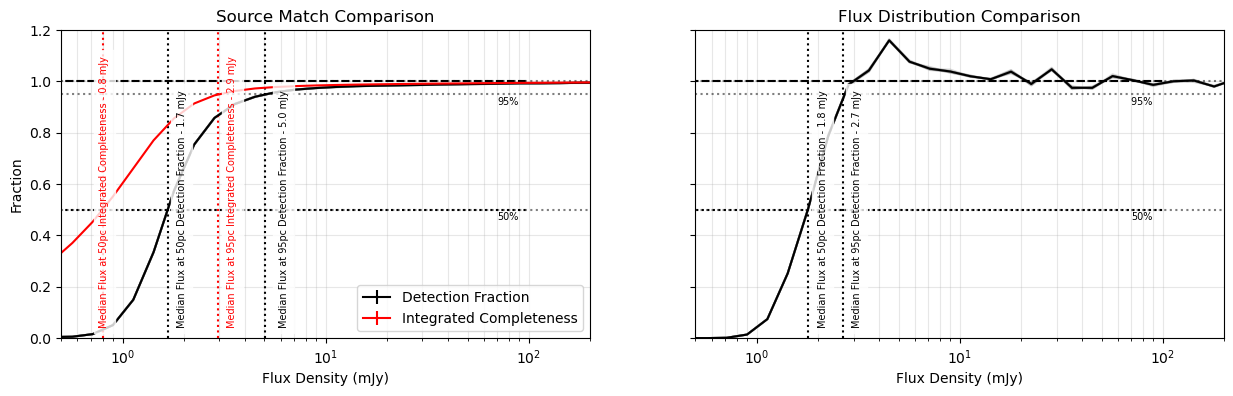}
\caption{{Detection fraction and completeness as a function of flux density for point source simulations. Left: Detection fraction and total catalogue completeness for sources which are matched based on positional location alone. The vertical lines indicate the 50\% and 95\% detection fraction levels (black; at 1.7 mJy and 5.0 mJy) and 50\% and 95\%  completeness (red; at 0.8 mJy and 2.9 mJy). Right: Detection fraction of sources as a function of flux density based on comparing the input to measured flux density distribution (50\% at 1.8 mJy and 95\% at 2.7 mJy). }}\label{fig:completeness}
\end{center}
\end{figure*}

First within each flux density bin, we investigate the fraction of sources that have a ``recovered" counterpart. The result of this can be seen in the left panel of Figure~\ref{fig:completeness}. This shows approximately 50\% detection at $\sim$1.7 mJy and 95\% detection at $\sim5.0$ mJy. From this, we also consider the overall completeness of the sources detected in the survey. To do this, we combine knowledge of the underlying flux distribution of sources from the SKADS simulations, with the fraction that are detected. For each flux density bin (logarithmically sampled) we sum the product of the detection fraction with the input source count distribution from the random sources at flux densities greater than or equal to the flux density bin being considered. This is normalised to the sum of the full input source count distribution. This overall completeness can be seen in the left hand of Figure~\ref{fig:completeness}. This suggests an overall 50\% completeness at $\sim$0.8 mJy and 95\% completeness at $\sim2.9$ mJy. 

\begin{figure*}[h!]
\begin{center}
\includegraphics[width=16cm]{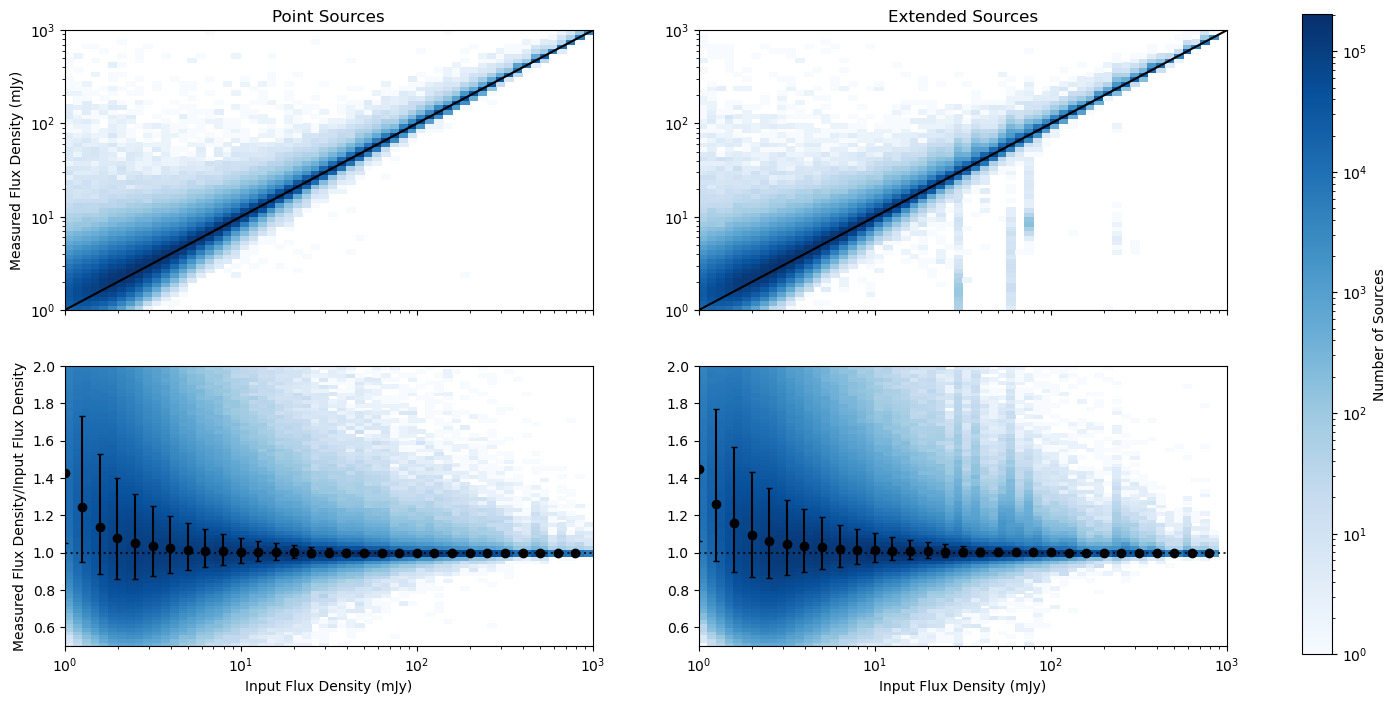}
\caption{{Comparison of input to measured fluxes for simulated sources in the (left) point source simulations and (right) resolved source simulations. Upper panels show the comparison of input and measured fluxes and lower panels show the median measured to input flux ratio as a function of input flux density. The black points indicate the median flux ratio within the bin and the errors are calculated from the 16$^{\textrm{th}}$ and 84$^{\textrm{th}}$ percentiles.}}
\label{fig:flux_sims_comparisons}
\end{center}
\end{figure*}

Secondly we consider the effect of flux measurement by the source finder and how this may affect the apparent distribution of fluxes. This comparison of input to measured flux distribution can be seen in Figure~\ref{fig:flux_sims_comparisons} (left panel for point sources). As can be seen, these measured fluxes are scattered around the 1-to-1 measurement line (black line), but will have a positive bias, especially at fainter flux densities. This positive bias is a combination of the effect of the measurement of source fluxes being affected by noise peaks/trough a source is on and, as brighter sources are more likely to be detected, sources which lie on a positive noise spike are more likely to be detected. Moreover, as there are more faint sources within the simulations, these are more likely to be affected by this positive bias.

To determine the point source detection fraction, with this second method, we compare the binned distribution of flux densities recovered by \texttt{PyBDSF} compared to the input flux density distribution of the simulated sources injected into the image. The ratio of the output flux density distribution compared to the input flux density distribution is therefore a measurement of the detection fraction of point sources across the image. This can be seen as the black line in the right panel of Figure~\ref{fig:completeness}. As the change in flux density can be seen through this measurement, it is possible to have detection fractions larger than one. This will reflect that, due to differences between input and output flux densities, there are more sources observed in a flux density bin than were input into the simulation. This method suggests a completeness of 50\% for point sources with a flux density of $\sim$1.8 mJy and 95\% at a flux density $\sim 2.7$ mJy. This method will be especially important in the discussion of source counts in Section~\ref{sec:sc}.

For both methods though, we determine the average detection fraction and completeness in each flux density bin by the mean value from the simulations. The associated error is then taken as the standard deviation from the simulation realisations.

\subsection{Resolved Source Completeness}
\label{sec:rescomp}
Next we investigate the effect of source size on completeness, as the previous section neglects the effect of resolution bias. Resolution bias accounts for the relative difficulty in detecting extended sources compared to point sources. An extended source with the same integrated flux density as a point source will have a lower peak flux density, an effect that becomes more important at low SNR. This will be important to consider as the majority of sources in this catalogue are believed to be resolved (see Figure~\ref{fig:envelope}). 

To investigate this, we again use the simulated sources from \cite{Wilman2008,Wilman2010}, using the source size models associated with each source. SKADS sources are described by a single or a combination of components which are described by ellipses. This will therefore contain a combination of single-component sources for objects such as SFGs as well as multi-component lobed FRI and FRII sources. These simulations should therefore give a more realistic distribution of the diverse ranges of sources expected within radio surveys and will contain a combination of resolved and unresolved sources.

\begin{figure*}[h!]
\begin{center}
\includegraphics[width=16cm]{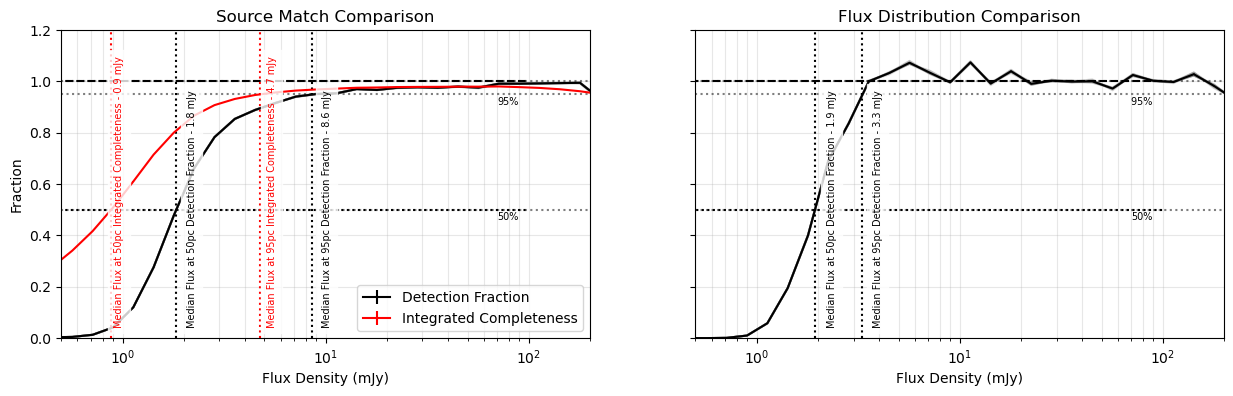}
\caption{{Detection fraction and completeness as a function of flux density for the resolved source simulations. Left: Detection fraction and total catalogue completeness for sources which are matched based on positional location alone. The vertical lines indicate the 50\% and 95\% detection fraction levels (black; at 1.8 mJy and 8.6 mJy) and 50\% and 95\%  completeness (red; at 0.9 mJy and 4.7 mJy). Right: Detection fraction of sources as a function of flux density based on comparing the input to measured flux density distribution (50\% at 1.9 mJy and 95\% at 3.3 mJy).}}\label{fig:completeness2}
\end{center}
\end{figure*}

To consider the completeness of a realistic distribution of resolved and unresolved sources, we follow the same method as in Section~\ref{sec:pointcomp} {however, first convolving the} ellipse source model with the Gaussian PSF, ensuring the flux scale is retained\footnote{{Some of the extremely large SKADS sources may have been truncated in injection into the image. However these sources were likely undetected in the image due to their peak fluxes and any contribution of these sources to the simulated sources are very small ($<<1$\%) and so should not largely affect results.}}. After running \texttt{PyBDSF} on each image\footnote{{We note that in pointings RACS\_1404-62A and RACS\_1314-62A, where there is significant extended emission within the image due to Galactic emission, most of the 10 simulations were unable to complete with reasonable computational tools in extended emission mode. We therefore ran these simulations with the extended atrous mode of \texttt{PyBDSF} switched off. These fields though are located close to the galactic plane and in fact after masking the galaxy for those regions with $|b|<5$\degree would not contribute to our catalogue. Therefore this should not affect the results.}}, we use the same process as in Section~\ref{sec:pointcomp} to {compare} the input and output catalogues and determine the detection fraction. This detection fraction is shown in Figure~\ref{fig:completeness2} and the comparison of input to measured flux densities can be seen in the right hand panel of Figure~\ref{fig:flux_sims_comparisons}. This shows {that} 50\% of sources at $\sim$1.8 (1.9) mJy will be detected, increasing to a 95\% detection fraction at approximately $\sim$8.6 (3.3) mJy using method 1 (2) described above. This indicates a poorer detection fraction than for point sources, reflecting the effect of resolution bias on the detection of sources. However, it could also relate to the fact that the simulated sources are extended and made of multiple components and so matching using a positional radius may lead to errors in matching components to a single source and so larger errors between the positional location of the input to measured source. {Issues due to this would also be seen in flux density comparisons in Figure \ref{fig:flux_sims_comparisons} where the input and measured flux densities appear offset.}. Computing the overall completeness of our catalogue from method 1 suggests 50\% completeness at $\sim$0.9 mJy and 95\% completeness at $\sim$4.7 mJy. 

\subsection{Limitations of the Simulations}
\label{sec:limitations}
We identify three separate limitations to the simulations we have used to analyse the survey completeness. First, these simulations inject sources into the residual image. Therefore these will not account for any issues that are introduced through the calibration pipeline or any effects of \texttt{CLEAN} bias \citep[see e.g. Section 7.2 of][]{FIRST} or time and bandwidth smearing \cite[see e.g.][]{BridleSchwab}. {Furthermore, some smearing may occur where images are mosaiced together which could affect source detectability, this includes both when beams are mosaiced together to form a tile and where tiles are mosaiced with other neighbouring tiles.} To improve this, sources could be injected into the visibilities and processed through the pipeline, however for \ntiles tiles, this is an arduous process. 

Second, these simulations will be affected if the source morphologies assumed in SKADS and flux density distribution of input sources are not as accurate a representation of the underlying source distribution as expected. This may also be the case if the morphologies observed with ASKAP are more susceptible to extended emission and complex morphologies which may not be well modelled in SKADS using elliptical components. In term of flux density distribution, though, the source counts from \cite{Wilman2008} seem to well recreate observations at the flux densities probed by RACS. This may, however, not be the case at fainter fluxes \citep[see e.g.][]{Smolcic2017, Mauch2020, Matthews2021}.

Finally, these simulations may not properly account for the effect of having multiple sources located in close proximity to each other. This is as the sources are injected at random positions and so will have a uniform source density. This will not account for the clustering of real sources due to the large scale structure of the Universe. Moreover, in the matching process there may be issues arising from simulated sources merging into a single source if they are located in close proximity. This could affect whether input to output simulated sources are matched together as well as any input to output flux ratios. 

Despite these limitations, these simulations should give a good understanding of how well we are detecting realistic radio sources within our images. We shall discuss further how successful these simulations appear to be, given their effect on the measured source counts, in Section~\ref{sec:sc}.

\subsection{Reliability}
Next we assess the reliability of these observations following the approach of \cite{TGSS}. This considers the contamination of noise within the catalogue by considering the source detection over the negative image (i.e. $-$1 $\times$ image). The technique relies on the premise that the noise in the image is symmetric. Therefore running \texttt{PyBDSF} over the negative images using the same parameters as in Section~\ref{sec:cat} can indicate the distribution of positive noise which may have contaminated the final source catalogue. 

\begin{figure*}
    \centering
    \includegraphics[width=12cm]{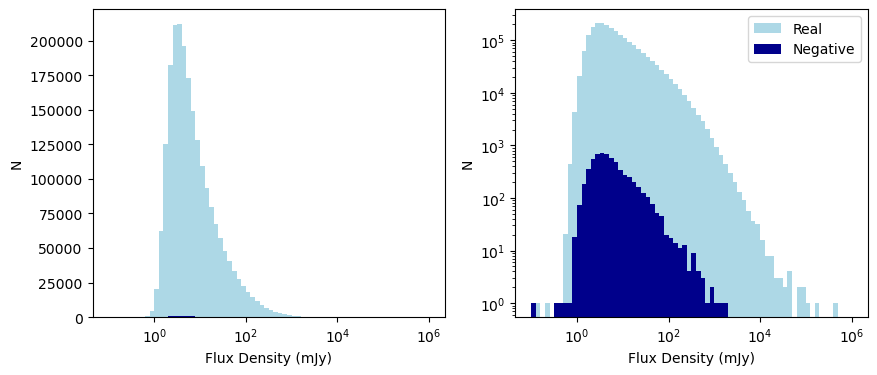}
    \caption{{Comparison of the flux density distribution of sources detected} in the negative image (dark blue) compared to the full survey catalogue (light blue). This is shown with both a linear (left) and logarithmic (right) scaled y-axis.}
    \label{fig:reliability}
\end{figure*}

We concatenate the \texttt{PyBDSF} catalogues from the negative images in the same way as described in Section~\ref{sec:cat}, to avoid source duplication. The distribution of source flux densities from the negative image compared to the final catalogues is shown in Figure~\ref{fig:reliability}. The number of negative sources is small compared to real sources within the catalogue ($\sim 0.3$\%), {suggesting that the number of false detections within the final source catalogue is negligible compared to real sources.}

\subsection{Density as function of declination}
\begin{figure*}[h!]
\begin{center}
\includegraphics[width=14cm]{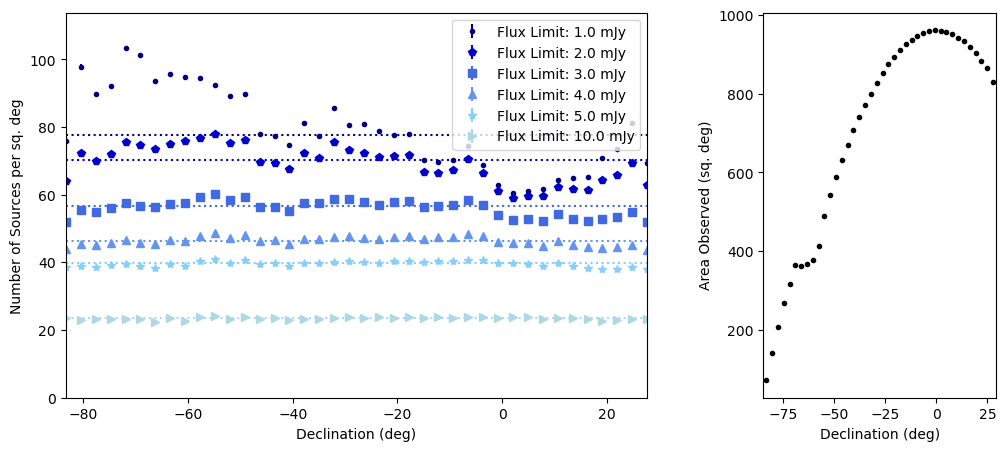}
\caption{{{Left: Source density as a function of Declination for those sources }with total flux densities above six limits. This is shown for the whole catalogue but with regions around the Galactic plane ($|b|<5$\degree) removed. Right: The corresponding sky area observed as a function of declination for the catalogue excluding the Galactic plane.}}\label{fig:source_density}
\end{center}
\end{figure*}

Finally, in order to consider the completeness as a function of sky position we present the variation of source density of the catalogue with declination. This is shown in Figure~\ref{fig:source_density} for all sources in the source catalogue, where we have excluded the Galactic plane. The density of sources with a total flux density above or equal to the six different flux density limits quoted (1, 2, 3, 4, 5 and 10 mJy) is shown. In Figure~\ref{fig:source_density}, the left hand panel shows this source density as a function of declination, whilst the right hand panel indicates the area which is being considered whilst constructing the source density. The integrated total number density of sources above a given flux density is presented in Table~\ref{tab:integratedsc}.

\begin{table}
\begin{center}
    \centering
\begin{tabular}{ccc}
$S$ & N ($>S$) & N ($>S$) \\
(mJy) & ($\rm{sr}^{-1}$) & ($\rm{deg}^{-2}$) \\ \hline \hline
0.5 & 249000.0 & 75.8 \\
1.0 & 248000.0 & 75.6 \\
2.0 & 224000.0 & 68.1 \\
3.0 & 183000.0 & 55.9 \\
5.0 & 130000.0 & 39.6 \\
10.0 & 77100.0 & 23.5 \\
20.0 & 43900.0 & 13.4 \\
50.0 & 19100.0 & 5.8 \\
100.0 & 9240.0 & 2.8 \\
200.0 & 3980.0 & 1.2 \\
\end{tabular}
\end{center}
\caption{{The integrated source counts of sources in the first RACS Stokes I catalogue above quoted flux density limits. We note that in the faintest flux density bins, the integral source counts will be affected by incompleteness \protect (see Section~\ref{sec:comp_rel}).}}
\label{tab:integratedsc}
\end{table}

As can be seen in Figure~\ref{fig:source_density}, the source density within our catalogue is approximately flat across the entire declination range observed for flux density limits $\geq 4$ mJy. In the 1 and 2 mJy flux density limit bins, on the other hand, the incompleteness limits within the data means that the source density is more variable over the declination range. These are still relatively small variations at 3 mJy; however, for the 1 mJy limit, the source density of sources around $\delta=-$60\degree \ is much larger compared to at other declination ranges. Moreover, due to the higher rms that can be seen in Figure~\ref{fig:rms_dist}, there is an under-density in sources at declinations of $\sim$0\degree \ to +20\degree \ in both the 1 and 2 mJy bins.

\begin{figure*}[h!]
\begin{center}
\includegraphics[width=17cm]{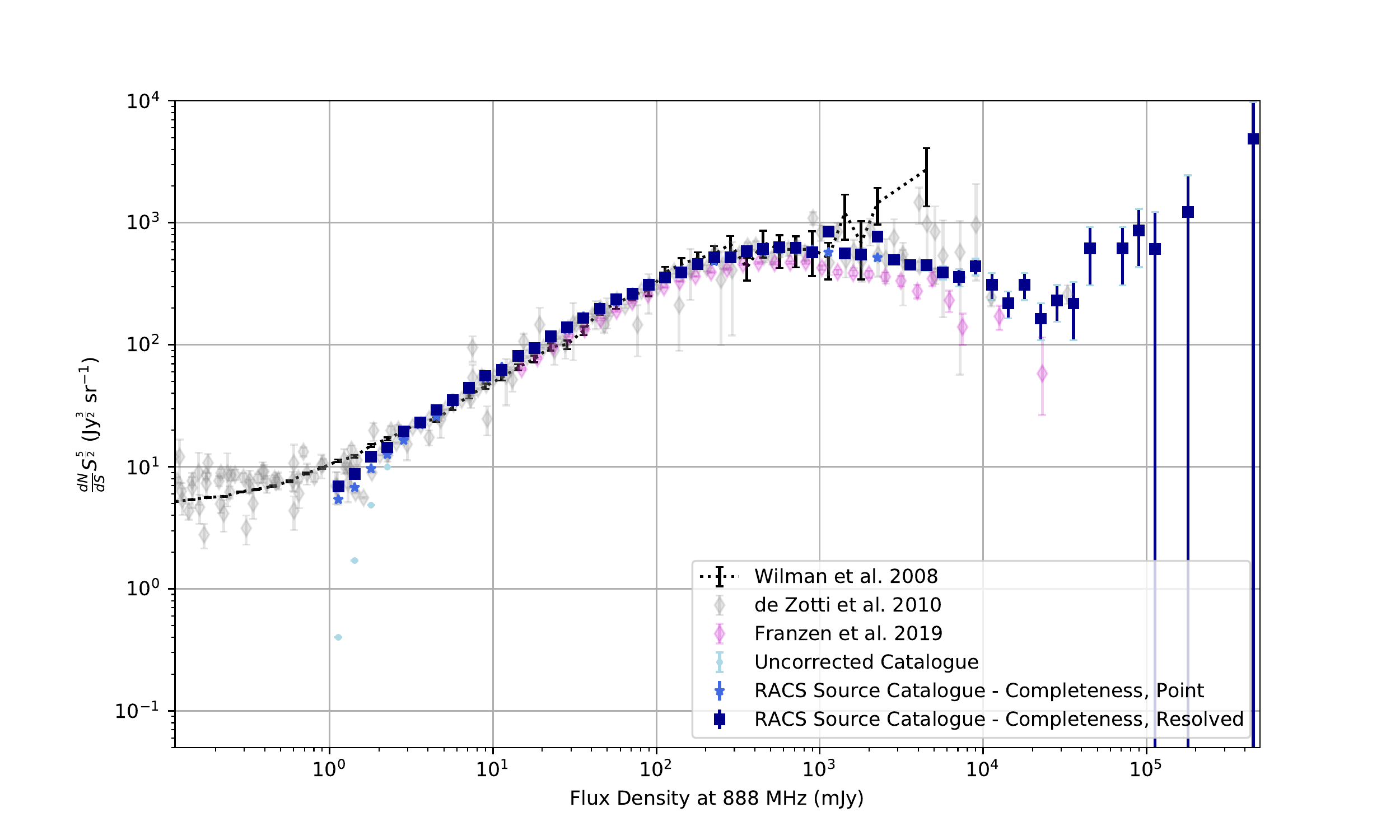}
\caption{{Comparison between the Euclidean normalised source counts} from this catalogue and previous surveys and simulations. Presented are the source counts from \protect \cite{deZotti2010} (grey diamonds), \protect \cite{Franzen2019} (magenta diamonds) and the SKADS catalogue from \cite{Wilman2008, Wilman2010} (black line). For our catalogue, we present the raw source counts (light blue) as well as those corrected using the point source detection fraction in Section~\ref{sec:pointcomp} (blue) and those using the variable size detection fraction from Section~\ref{sec:rescomp} (dark blue). The RACS number counts cover an area of \skyareagalcut deg$^2$. All surveys not at 887.5 MHz have been converted to this frequency assuming $\alpha=-0.8$. }
\label{fig:SourceCounts}
\end{center}
\end{figure*}

\section{Source Counts}
\label{sec:sc}
Finally, using our finished catalogue and, having quantified the completeness within our sample, we compare the source count distribution of the radio sources identified in our catalogue to previous surveys. Whilst narrow, deep surveys help to fill in the source count distribution of faint sources, it is only with large area surveys that the source count distribution of the brightest sources can be understood. This is because these bright objects are rare. Source counts describe the number of sources within a given flux density bin per steradian on the sky. These are typically normalised by multiplying by $S^{2.5}$ (where $S$ is the total flux density) to define the Euclidean normalised source counts \citep[see e.g.][for an explanation]{Heywood2013}.

Radio source counts are constructed for most radio survey catalogues \citep[see e.g.][]{Bondi2008, Smolcic2017, LoTSS} and so compilations of the source counts from multiple surveys exist \citep[e.g.][]{deZotti2010}. We therefore determine the source counts for our catalogue and make comparisons to the past survey source counts compiled by \cite{deZotti2010} and the low frequency source counts from GLEAM from \cite{Franzen2019}, which cover large sky areas, in Figure~\ref{fig:SourceCounts}. Here we illustrate the source count distribution for the source catalogue discussed in Section~\ref{sec:cat}. The raw count from this catalogue can be seen as the light blue circles in Figure~\ref{fig:SourceCounts}.

However, as discussed in Section~\ref{sec:comp_rel}, our observations are not complete to the faintest flux density limits {to which we observe}. This is due to the fact that there are noise variations across the full survey area of RACS, meaning that the faintest sources are unable to be detected uniformly. To make a correction for this and to correct the measured source counts to what would be observed if the rms was uniform, we make use of the detection fraction curves described in Section~\ref{sec:comp_rel}. Specifically, we use the detection fraction where we account for variations in flux density (see Figures~\ref{fig:completeness} and~\ref{fig:completeness2} - right). We use this as the measurements of the flux densities of sources will also be affected by any differences between the true and measured source densities.

Using the detection fraction as a function of flux density, we corrected the raw counts using the detection fractions from Sections~\ref{sec:pointcomp} and~\ref{sec:rescomp} to understand the {intrinsic} source count distribution. We calculated the associated errors by adding in quadrature the errors from both the poisson statistics of the data itself, with the errors from the completeness simulations (see Section~\ref{sec:pointcomp}). In Figure~\ref{fig:SourceCounts}, we plot the source counts only for those sources that have a flux density (at 887.5 MHz) of greater than 1 mJy. These are compared to the compilation of measured source counts from \cite{deZotti2010} as well as the source counts from the extra-galactic simulated catalogues of SKADS (both converted to 887.5 MHz assuming $\alpha=-$0.8). We apply corrections based on both the point source only corrected simulations (Section~\ref{sec:pointcomp}; blue) as well as the simulations which have both point and extended simulated sources (Section~\ref{sec:rescomp}; dark blue). These are both plotted so that the effect of source size can be investigated.

As can be seen from Figure~\ref{fig:SourceCounts}, these corrections only affect the lowest flux density bins, below approximately 5 mJy. Using the corrections from Section~\ref{sec:comp_rel} we find that if we only include point sources in our investigations, then the source counts appear too small at faint flux densities in comparison to previous observations. When the effect of extended sources is included, these source counts are {further corrected} and are now in much better agreement with the source counts from \cite{deZotti2010} and \cite{Wilman2008}. However, these source counts may still possibly appear too low in the faintest flux density bins and possible explanations for this can be found in Section~\ref{sec:limitations}. A table of the resulting RACS source counts can be found in Table~\ref{tab:tab_sc}.

At high flux densities, we find that RACS is able to provide tight constraints on the source counts, due to the large area coverage of the survey. Importantly, this allows the source counts at flux densities at values $\gtrsim$ 10$^4$ mJy that are not well investigated in the \cite{deZotti2010} compilation catalogue to be seen. The source counts presented at these high flux densities do appear higher than the counts from \cite{Franzen2019}. This may reflect differences in the source populations observed at lower frequencies (200 MHz).

\begin{table*}[]
    \centering
    \begin{tabular}{r r r r r r }
$S$ & $S_{mid}$ & N & Raw $\frac{dN}{dS} S^{2.5}$ & Corrected $\frac{dN}{dS} S^{2.5}$  & Corrected $\frac{dN}{dS} S^{2.5}$  \\ 
& & & & (Point) & (Resolved) \\
(mJy) & (mJy) &  & (Jy$^{1.5}$sr$^{-1}$) &(Jy$^{1.5}$sr$^{-1}$) & (Jy$^{1.5}$sr$^{-1}$) \\ \hline \hline
1.00 - 1.26 & 1.13 & 20653 $\pm$ 143 & 0.40 $\pm$ 0.01 & 5.41 $\pm$ 0.04 & 6.91 $\pm$ 0.05 \\ 
1.26 - 1.58 & 1.42 & 62252 $\pm$ 249 & 1.71 $\pm$ 0.01 & 6.77 $\pm$ 0.03 & 8.76 $\pm$ 0.03 \\ 
1.58 - 2.00 & 1.79 & 125409 $\pm$ 354 & 4.85 $\pm$ 0.01 & 9.67 $\pm$ 0.04 & 12.14 $\pm$ 0.05 \\ 
2.00 - 2.51 & 2.25 & 182302 $\pm$ 426 & 9.97 $\pm$ 0.02 & 12.64 $\pm$ 0.05 & 14.42 $\pm$ 0.04 \\ 
2.51 - 3.16 & 2.84 & 211331 $\pm$ 459 & 16.32 $\pm$ 0.04 & 16.52 $\pm$ 0.05 & 19.52 $\pm$ 0.05 \\ 
3.16 - 3.98 & 3.57 & 212212 $\pm$ 460 & 23.15 $\pm$ 0.05 & 22.20 $\pm$ 0.08 & 23.13 $\pm$ 0.06 \\ 
3.98 - 5.01 & 4.50 & 196155 $\pm$ 442 & 30.22 $\pm$ 0.07 & 26.05 $\pm$ 0.07 & 29.26 $\pm$ 0.08 \\ 
5.01 - 6.31 & 5.66 & 172900 $\pm$ 415 & 37.63 $\pm$ 0.09 & 34.93 $\pm$ 0.11 & 35.08 $\pm$ 0.16 \\ 
6.31 - 7.94 & 7.13 & 149416 $\pm$ 386 & 45.93 $\pm$ 0.12 & 43.76 $\pm$ 0.18 & 44.42 $\pm$ 0.20 \\ 
7.94 - 10.00 & 8.97 & 127989 $\pm$ 357 & 55.58 $\pm$ 0.16 & 53.50 $\pm$ 0.22 & 55.76 $\pm$ 0.17 \\ 
10.00 - 12.59 & 11.29 & 109141 $\pm$ 330 & 66.94 $\pm$ 0.20 & 65.61 $\pm$ 0.23 & 62.32 $\pm$ 0.26 \\ 
12.59 - 15.85 & 14.22 & 93233 $\pm$ 305 & 80.78 $\pm$ 0.26 & 80.08 $\pm$ 0.29 & 81.44 $\pm$ 0.35 \\ 
15.85 - 19.95 & 17.90 & 79828 $\pm$ 282 & 97.69 $\pm$ 0.35 & 94.11 $\pm$ 0.48 & 93.96 $\pm$ 0.47 \\ 
19.95 - 25.12 & 22.54 & 67405 $\pm$ 259 & 116.52 $\pm$ 0.45 & 117.65 $\pm$ 0.62 & 117.59 $\pm$ 0.64 \\ 
25.12 - 31.62 & 28.37 & 57111 $\pm$ 238 & 139.46 $\pm$ 0.58 & 133.16 $\pm$ 0.67 & 139.05 $\pm$ 0.64 \\ 
31.62 - 39.81 & 35.72 & 48027 $\pm$ 219 & 165.65 $\pm$ 0.76 & 169.85 $\pm$ 0.97 & 165.73 $\pm$ 0.88 \\ 
39.81 - 50.12 & 44.96 & 40703 $\pm$ 201 & 198.31 $\pm$ 0.98 & 203.36 $\pm$ 1.16 & 198.17 $\pm$ 1.26 \\ 
50.12 - 63.10 & 56.61 & 33434 $\pm$ 182 & 230.09 $\pm$ 1.25 & 225.41 $\pm$ 1.41 & 236.64 $\pm$ 1.41 \\ 
63.10 - 79.43 & 71.26 & 27623 $\pm$ 166 & 268.53 $\pm$ 1.61 & 267.23 $\pm$ 1.66 & 261.96 $\pm$ 1.78 \\ 
79.43 - 100.00 & 89.72 & 22715 $\pm$ 150 & 311.91 $\pm$ 2.06 & 316.13 $\pm$ 2.20 & 311.06 $\pm$ 2.15 \\ 
100.00 - 125.89 & 112.95 & 18415 $\pm$ 135 & 357.18 $\pm$ 2.62 & 356.92 $\pm$ 2.71 & 357.94 $\pm$ 2.72 \\ 
125.89 - 158.49 & 142.19 & 14763 $\pm$ 121 & 404.47 $\pm$ 3.32 & 402.96 $\pm$ 3.45 & 393.35 $\pm$ 3.84 \\ 
158.49 - 199.53 & 179.01 & 11648 $\pm$ 107 & 450.78 $\pm$ 4.14 & 459.76 $\pm$ 4.28 & 458.81 $\pm$ 4.72 \\ 
199.53 - 251.19 & 225.36 & 8908 $\pm$ 94 & 486.96 $\pm$ 5.14 & 483.41 $\pm$ 5.16 & 523.56 $\pm$ 5.45 \\ 
251.19 - 316.23 & 283.71 & 6898 $\pm$ 83 & 532.64 $\pm$ 6.41 & 533.12 $\pm$ 6.49 & 522.06 $\pm$ 6.85 \\ 
316.23 - 398.11 & 357.17 & 5228 $\pm$ 72 & 570.23 $\pm$ 7.85 & 580.13 $\pm$ 8.44 & 584.12 $\pm$ 8.72 \\ 
398.11 - 501.19 & 449.65 & 3815 $\pm$ 61 & 587.77 $\pm$ 9.40 & 581.66 $\pm$ 9.40 & 610.97 $\pm$ 11.54 \\ 
501.19 - 630.96 & 566.07 & 2859 $\pm$ 53 & 622.20 $\pm$ 11.53 & 606.49 $\pm$ 11.46 & 629.78 $\pm$ 11.46 \\ 
630.96 - 794.33 & 712.64 & 2047 $\pm$ 45 & 629.26 $\pm$ 13.83 & 631.10 $\pm$ 13.97 & 622.63 $\pm$ 14.46 \\ 
794.33 - 1000.00 & 897.16 & 1389 $\pm$ 37 & 603.14 $\pm$ 16.07 & 600.81 $\pm$ 16.14 & 572.48 $\pm$ 17.90 \\ 
1000.00 - 1258.93 & 1129.46 & 931 $\pm$ 30 & 571.04 $\pm$ 18.40 & 574.04 $\pm$ 18.62 & 850.87 $\pm$ 21.91 \\ 
1258.93 - 1584.89 & 1421.91 & 644 $\pm$ 25 & 557.96 $\pm$ 21.66 & 556.81 $\pm$ 21.65 & 559.08 $\pm$ 21.70 \\ 
1584.89 - 1995.26 & 1790.08 & 448 $\pm$ 21 & 548.27 $\pm$ 25.70 & 547.25 $\pm$ 25.70 & 549.16 $\pm$ 25.69 \\ 
1995.26 - 2511.89 & 2253.57 & 300 $\pm$ 17 & 518.60 $\pm$ 29.39 & 518.29 $\pm$ 29.39 & 768.21 $\pm$ 31.82 \\ 
2511.89 - 3162.28 & 2837.08 & 205 $\pm$ 14 & 500.57 $\pm$ 34.19 & 500.57 $\pm$ 34.19 & 497.07 $\pm$ 34.41 \\ 
3162.28 - 3981.07 & 3571.67 & 131 $\pm$ 11 & 451.84 $\pm$ 37.94 & 452.25 $\pm$ 38.00 & 451.84 $\pm$ 37.98 \\ 
3981.07 - 5011.87 & 4496.47 & 92 $\pm$ 9 & 448.23 $\pm$ 43.85 & 450.67 $\pm$ 44.12 & 450.99 $\pm$ 44.16 \\ 
5011.87 - 6309.57 & 5660.72 & 57 $\pm$ 7 & 392.27 $\pm$ 48.17 & 387.84 $\pm$ 47.71 & 392.27 $\pm$ 47.63 \\ 
6309.57 - 7943.28 & 7126.43 & 37 $\pm$ 6 & 359.68 $\pm$ 58.33 & 359.68 $\pm$ 58.33 & 359.68 $\pm$ 58.33 \\ 
7943.28 - 10000.00 & 8971.64 & 32 $\pm$ 5 & 439.40 $\pm$ 68.66 & 439.40 $\pm$ 68.66 & 439.40 $\pm$ 68.66 \\ 
10000.00 - 12589.25 & 11294.63 & 16 $\pm$ 4 & 310.34 $\pm$ 77.58 & 310.34 $\pm$ 77.58 & 310.34 $\pm$ 77.58 \\ 
12589.25 - 15848.93 & 14219.09 & 8 $\pm$ 2 & 219.18 $\pm$ 54.80 & 219.18 $\pm$ 54.80 & 219.18 $\pm$ 54.80 \\ 
15848.93 - 19952.62 & 17900.78 & 8 $\pm$ 2 & 309.60 $\pm$ 77.40 & 309.60 $\pm$ 77.40 & 309.60 $\pm$ 77.40 \\ 
19952.62 - 25118.86 & 22535.74 & 3 $\pm$ 1 & 164.00 $\pm$ 54.67 & 164.00 $\pm$ 54.67 & 164.00 $\pm$ 54.67 \\ 
25118.86 - 31622.78 & 28370.82 & 3 $\pm$ 1 & 231.65 $\pm$ 77.22 & 231.65 $\pm$ 77.22 & 231.65 $\pm$ 77.22 \\ 
31622.78 - 39810.72 & 35716.75 & 2 $\pm$ 1 & 218.14 $\pm$ 109.07 & 218.14 $\pm$ 109.07 & 218.14 $\pm$ 109.07 \\ 
39810.72 - 50118.72 & 44964.72 & 4 $\pm$ 2 & 616.27 $\pm$ 308.14 & 616.27 $\pm$ 308.14 & 616.27 $\pm$ 308.14 \\ 
63095.73 - 79432.82 & 71264.28 & 2 $\pm$ 1 & 614.82 $\pm$ 307.41 & 614.82 $\pm$ 307.41 & 614.82 $\pm$ 307.41 \\ 
79432.82 - 100000.00 & 89716.41 & 2 $\pm$ 1 & 868.45 $\pm$ 434.22 & 868.45 $\pm$ 434.22 & 868.45 $\pm$ 434.22 \\ 
100000.00 - 125892.54 & 112946.27 & 1 $\pm$ 1 & 613.36 $\pm$ 613.36 & 613.36 $\pm$ 613.36 & 613.36 $\pm$ 613.36 \\ 
158489.32 - 199526.23 & 179007.78 & 1 $\pm$ 1 & 1223.81 $\pm$ 1223.81 & 1223.81 $\pm$ 1223.81 & 1223.81 $\pm$ 1223.81 \\ 
398107.17 - 501187.23 & 449647.20 & 1 $\pm$ 1 & 4872.08 $\pm$ 4872.08 & 4872.08 $\pm$ 4872.08 & 4872.08 $\pm$ 4872.08 \\ 
   \end{tabular}
    \caption{{Table of source counts at 887.5 MHz accompanying Figure \protect~\ref{fig:SourceCounts}. The raw and the corrected source counts using the corrections for point (Section~\ref{sec:pointcomp}) and resolved (Section~\ref{sec:rescomp}) are quoted. A total sky area of \skyareagalcut deg$^2$ was used.}}
    \label{tab:tab_sc}
\end{table*}

\section{Conclusions}
\label{sec:conclusions}
In this paper we have presented the first Stokes I catalogue release for the Rapid ASKAP Continuum Survey \citep[RACS,][]{RACS}. This first catalogue contains the majority of the Southern sky in the Declination region $-$80\degree \ to +30\degree \ using \ntiles tiles across the sky. These observations were reduced as described in \cite{RACS} and in this paper we describe the process of mosaicing the observations together and producing a single source catalogue. We present the source catalogue of {\nislgalcut}  \ islands and {\ncompgalcut}  \ components which were derived from \texttt{PyBDSF} catalogues over each of the \ntiles tiles. These have been combined together to form a full catalogue which removes duplicate sources. This catalogue will be released for download from CASDA {\citep{CASDA, CASDA2}} with the publication of the paper.

For quality assessment, we have compared the results of this work to previous large sky radio surveys from GLEAM, NVSS, SUMSS and TGSS-ADR. This has allowed quantification of the accuracy in the flux density scale and astrometry, along with the spectral indices implied from our data. We find good flux density agreement with SUMSS, finding a RACS-to-SUMSS flux density ratio of $\flsumss^{+\flsumssperr}_{-\flsumssnerr}$. Our median astrometric offsets from comparisons with SUMSS and NVSS appear to be limited to a pixel (2.5\arcsec) with most offsets constrained to less than two pixels. Finally, we find typical $\alpha$ measurements of $\sim -$0.6 compared to radio observations at lower frequency to RACS and $\sim -$0.9 for surveys at higher frequencies. 

We have further analysed the data using simulations to investigate the detection fraction of both point and resolved sources within our images as a function of flux density. Using these measurements, we determined that this catalogue detects 95\% of point sources at $\sim 5$ mJy, leading to a 95\% total point source completeness at $\sim$3 mJy. We have shown using the detection fraction of sources (of varying size) that we can recover source count distributions similar to previous work over the range $1-10^4$ mJy, and we improve knowledge of the source count distribution at the highest flux densities ($\gtrsim10^4$ mJy) compared to previous work. 

In summary, this work has described the first large sky RACS catalogue, which provides the deepest radio observations of the southern sky to date. This is especially impressive given the brief duration of each observation and the short overall survey time. We have constructed a deep catalogue of the radio sky at 887.5 MHz, which is important for radio sky models and science. In the future we will improve upon this first catalogue, providing further catalogues to fill in gaps over the southern sky as well as catalogues at other frequencies. This will provide more information about the spectral distribution of sources within the southern sky. Moreover, the Stokes Q and U polarisation products from RACS will be used to produce a corresponding linear polarization catalogue, known as SPICE-RACS (Spectra and Polarization In Cutouts of Extragalactic Sources from RACS, Thomson et al. in prep).

\begin{acknowledgements}
{We thank the referee for their helpful comments to improve this manuscript and we also thank Minh Huynh for help  in uploading the data to the CSIRO ASKAP Science Data Archive, CASDA}. The Australian SKA Pathfinder is part of the Australia Telescope National Facility which is managed by CSIRO. Operation of ASKAP is funded by the Australian Government with support from the National Collaborative Research Infrastructure Strategy. ASKAP uses the resources of the Pawsey Supercomputing Centre. Establishment of ASKAP, the Murchison Radio-astronomy Observatory and the Pawsey Supercomputing Centre are initiatives of the Australian Government, with support from the Government of Western Australia and the Science and Industry Endowment Fund. We acknowledge the Wajarri Yamatji people as the traditional owners of the Observatory site. This work was supported by resources provided by the Pawsey Supercomputing Centre with funding from the Australian Government and the Government of Western Australia. CLH acknowledges support from the Leverhulme Trust through an early career research fellowship. TM acknowledges the support of the Australian Research Council through grants FT150100099 and DP190100561. JL  and JP are supported by Australian Government Research Training Program Scholarships. 

Results in this paper have been derived using several packages: healpy and HEALPix package \citep{Gorski2005, Zonca2019}, Astropy \citep{Astropy}, Scipy \citep{scipy}, Numpy \citep{numpy}, matplotlib \citep{Matplotlib}, tqdm (\url{https://doi.org/10.5281/zenodo.4586769}). This research made use of APLpy, an open-source plotting package for Python \citep{aplpy1,aplpy2}. We have also made use of programs such as ds9 \citep{ds9}, TOPCAT \citep{topcat} and Aladin \citep{aladin} in order to help with this research. This publication makes use of data products from the Wide-field Infrared Survey Explorer, which is a joint project of the University of California, Los Angeles, and the Jet Propulsion Laboratory/California Institute of Technology, funded by the National Aeronautics and Space Administration.

\end{acknowledgements}

\bibliographystyle{pasa-mnras}
\bibliography{bibtex}

\end{document}